\documentclass[journal=jpccck, manuscript=article]{achemso}
\setkeys{acs}{articletitle = true}
\usepackage{graphics,dcolumn}
\usepackage{graphicx}
\usepackage{amsmath}
\usepackage{amsfonts}
\usepackage{pgf}
\usepackage{soul}
\SectionNumbersOn

\title{
{\footnotesize This document is the unedited author’s version of a
Submitted Work that was subsequently accepted for
publication in The Journal of Physical Chemistry C,
copyright
\textsuperscript{\textcopyright}ACS after
peer review. To access the final edited and published work see:
https://pubs.acs.org/doi/abs/10.1021/acs.jpcc.8b12319
}\\
Driven Liouville-von Neumann Equation for Quantum
Transport and Multiple-Probe Green's Functions}

\author{Francisco Ram\'irez}
\affiliation{Departamento de Qu\'imica Inorg\'anica,
Anal\'itica y Qu\'imica F\'isica/INQUIMAE, Facultad
de Ciencias Exactas y Naturales, Universidad de Buenos
Aires, Ciudad Universitaria, Buenos Aires (C1428EHA)
Argentina}
\email{framirez@qi.fcen.uba.ar}

\author{Daniel Dundas}
\affiliation{Atomistic Simulation Centre, School of
Mathematics and Physics, Queen's University Belfast,
Belfast BT7 1NN, UK}

\author{Cristi\'an G. S\'anchez}
\affiliation{Departamento de Qu\'imica Te\'orica y
Computacional, Facultad de Ciencias Qu\'imicas,
Universidad Nacional de C\'ordoba, Ciudad Universitaria
X5000HUA, C\'ordoba Argentina, and CONICET \& Facultad
de Ciencias Exactas y Naturales, Universidad Nacional
de Cuyo, Mendoza, CP5500, Argentina}

\author{Damian A. Scherlis}
\affiliation{Departamento de Qu\'imica Inorg\'anica,
Anal\'itica y Qu\'imica F\'isica/INQUIMAE, Facultad
de Ciencias Exactas y Naturales, Universidad de Buenos
Aires, Ciudad Universitaria, Buenos Aires (C1428EHA)
Argentina}

\author{Tchavdar N. Todorov}
\affiliation{Atomistic Simulation Centre, School of
Mathematics and Physics, Queen's University Belfast,
Belfast BT7 1NN, UK}

\date{\parbox{\linewidth}{\centering%
  \today\endgraf\bigskip
  Coordinator 1 \hspace*{3cm} Coordinator 2\endgraf\medskip
  Dept.\ of Physics \endgraf
  ABC Colleg
}}

\begin{document}

\begin{abstract}
The so called Driven Liouville-von Neumann equation is a
dynamical formulation to simulate a voltage bias across a
molecular system and to model a time-dependent current
in a grand-canonical framework.
This approach introduces a damping term in the equation
of motion that drives the charge to a reference, out of equilibrium density.
Originally proposed by Horsfield and co-workers, further
work on this scheme has led to different coexisting
versions of this equation.
On the other hand, the multiple-probe scheme devised by
Todorov and collaborators, known as the hairy-probes
method, is a formal treatment based on Green's functions
that allows to fix the electrochemical potentials in two
regions of an open quantum system.
In this article, the equations of motion of the hairy
probes formalism are rewritten to show that, under certain
conditions, they can assume the same algebraic structure as
the Driven Liouville-von Neumann equation in the form proposed by Morzan et al. 
[J. Chem. Phys. {\bf 2017}, 146, 044110].
In this way, a new formal ground is provided for the latter,
identifying  the origin of every term.
The performance of the different methods are explored
using tight-binding time-dependent simulations in three
trial structures, designated as ballistic, disordered,
and resonant models.
In the context of first-principles Hamiltonians the Driven
Liouville-von Neumann approach is of special interest, 
because it does not require the calculation of Green's
functions.
Hence, the effects of replacing the reference density based
on the Green's function by one obtained from an applied
field are investigated, to gain a deeper understanding of
the limitations and the range of applicability of the Driven
Liouville-von Neumann equation.
\end{abstract}

\section{Introduction}
\label{section.intro}

The interest in molecular conductance and electronic
transport across nanostructures has inspired the
development of a multiplicity of theoretical treatments
to compute the current under an applied bias.
The proposed schemes vary in complexity and computational
cost, going from the original Landauer-B\"{u}ttiker method
\cite{landauer1, buttiker} and different static models
meant to describe steady state transport,
\cite{PRB-Transiesta, ref7-jcp_120_7165,
ref8-jcp_126_144104, jpcm_20_083203, jcp_148_030901}
to dynamical methodologies that take into account the
time evolution of the charge density.
\cite{ref09-PhysRevB.72.035308, PRL-Gebauer-Burke,
hp, ref10-PhysRevLett.100.176403, jcp_148_030901,
jpcm_20_083203} 
The present article is concerned with the later class of
approaches, and in particular with the subset of methods
based on the so-called Driven Liouville-von Neumann (DLvN)
equation.
In this framework, the dynamics at the electrodes is
modulated by an additional driving term that augments the
standard Liouville-von Neumann equation of motion.
The role of these terms is to enforce part of the density
matrix associated with the electrodes to remain close to a
reference density matrix, thus introducing a charge
imbalance between a ``source'' or left lead ($L$) and a
``drain'' or right lead ($R$).
This driving term originally takes the form of a
difference between the time-dependent density matrix
$\hat \rho (t)$, and a reference density matrix
$\hat \rho_0$,
\begin{equation}
 \dot{\hat{\rho}}(t) = -\frac{i}{\hbar} [ \hat{H}, \hat{\rho}(t)]  
- \Gamma (\hat{\rho}(t) - \hat{\rho}^0)
\end{equation}
where $\hat H$ is the electron Hamiltonian, 
$\Gamma$ is the driving rate parameter, and the matrix $\hat \rho^0$ can be defined as follows:
\begin{equation}
  {\rho}_{ij}^0 = \left\{
        \begin{array}{ll}
            \rho_{ij}(t_0) & \text{if  $i,j$ $\in$ $L \cup R$ } \\
            \rho_{ij}(t) & \text{if $i,j$ $\notin$ $L \cup R$ } \\
        \end{array}
    \right.
\end{equation}

This type of approach was first introduced by Horsfield
and co-workers \cite{jpcm_16_L65, jcp_124_214708}
as an intuitive way of including at the level of the
density matrix the circulation of charge between the
electrodes, and it was further enriched by the work
of Nitzan \cite{jcp_130_144105} and Mazziotti
\cite{jcp_132_104112}.
These ideas were later taken on by Hod and co-workers \cite{jctc_10_2927,ML-Coup-Hod-JCTC,linblad_hod,
nonorthogonal} and reelaborated by 
deriving the driving terms from the theory of complex
absorbing potentials. In doing so, they arrived at a
modified form in which coherences were damped to zero,
\begin{equation}
 \dot{\hat{\rho}}(t)= -\frac{i}{\hbar} [ \hat{H}, \hat{\rho}(t)]
-\frac{\Gamma}{2}
\begin{bmatrix}
2(\hat{\rho}_{LL} - \hat{\rho}_{LL}^0) & \hat{\rho}_{LC}  & 2\hat{\rho}_{LR} \\
\hat{\rho}_{CL}   & 0  & \hat{\rho}_{CR}  \\
2\hat{\rho}_{RL}  & \hat{\rho}_{RC}  & 2(\hat{\rho}_{RR} - \hat{\rho}_{RR}^0)  \\
 \end{bmatrix}  ,
\label{master-tddft-z}
\end{equation}
that improved the stability of the calculations and the
steady-state convergence, and that was shown to satisfy
Pauli's principle regardless of the initial conditions.
\cite{jctc_10_2927}
Franco and collaborators presented a formal derivation
of this formula from the theory of non-equilibrium
Green's functions, demonstrating  that it can accurately
capture time-dependent transport phenomena.\cite{franco}
More recently, Zelovich et al. proposed a strategy to replace
the single rate parameter $\Gamma$ by diagonal matrices containing
the broadening factors corresponding to the lead states.\cite{parameterfree-DLvN}
These factors can be computed from the
self-energies of the electrodes, thus providing a parameter-free version of the Driven Liouville-von Neumann
approach.\cite{parameterfree-DLvN}

Because of its conceptual simplicity and good compromise
between computational cost and physical accuracy, the DLvN
method has attracted significant attention, and several
further refinements of its implementation and analysis of
its theoretical foundation were made.
Among these, of particular relevance is the adaptation to
a first-principles real-time TDDFT framework carried out
by Morzan and co-authors,\cite{jcp_morzan} where an
observed imbalance between injection and absorption of
charge during the dynamics prompted a reformulation of
the driving term. In an orthonormal basis, this equation
of motion assumes the following structure:

\begin{equation}
 \dot{\hat{\rho}}(t)= -\frac{i}{\hbar} [ \hat{H}, \hat{\rho}(t)]
-\frac{\Gamma}{2}
\begin{bmatrix}
2(\hat{\rho}_{LL} - \hat{\rho}_{LL}^0) &
\hat{\rho}_{LC} - \hat{\rho}_{LC}^0 &
2(\hat{\rho}_{LR} - \hat{\rho}_{LR}^0) \\
\hat{\rho}_{CL} - \hat{\rho}_{CL}^0  &
0  &
\hat{\rho}_{CR} - \hat{\rho}_{CR}^0 \\
2(\hat{\rho}_{RL} - \hat{\rho}_{RL}^0) &
\hat{\rho}_{RC} - \hat{\rho}_{RC}^0 &
2(\hat{\rho}_{RR} - \hat{\rho}_{RR}^0 ) \\
 \end{bmatrix}
\label{master-tddft-e}
\end{equation}
where the subscripts indicate the corresponding blocks of
the time-dependent density matrix, and $\hat{\rho}^0$
denotes a reference, time-independent density matrix.
A major change with respect to previous expressions
is that here all off-diagonal contributions are damped
to their reference values (obtained e.g. from the
polarized density matrix).
This proved to be important for charge conservation and
an appropriate balance between incoming and outgoing
currents in the relatively small models tractable in
TDDFT simulations.\cite{jcp_morzan}
Despite intuitively justifying this modification both
from a numerical standpoint and interpreting it as a
change in the boundary conditions, no formal theoretical
derivation was provided at the time.

In parallel with these developments, a different
driving term for the Liouville-von Neumann equation
was put forward by Todorov and co-workers.\cite{hp}
This new embedding method called multiple-probe or
``hairy probes'' (HP), formally derived a different
structure of the driving term via the application of
Green's functions and the Lippmann-Schwinger equation.
\cite{hp, ref22-todorov2002tight}
In spite of  its good results and wide range of
applicability,\cite{ref23-dundas2009current,
ref24-mceniry2010modelling,ref25-horsfield2016efficient}
its relationship with the reference density driving
method remained undetermined.

The objective of the present study is to bridge the gap
between the two methodologies: the heuristic formulation
of the Driven Liouville-von Neumann equation, and the
hairy-probes scheme.
For that purpose, we have been able to rewrite the
driving terms of the HP theory into two terms, one
of which resembles the damping term of the reference
density approach.
In doing so we have found that the latest addition of
Morzan and co-workers is the most compatible with HP,
thus finally providing a formal framework for their
correction of the non-diagonal blocks of the driving
term. 
Through the application of these schemes to a series
of model systems, we better characterize these
methodologies for different situations and shed
light on the reliability of the equation of motion
including the driving term with the reference
density matrix $\hat{\rho}^0$.

In the next two sections we introduce the theoretical
and methodological framework for this work: we first
show how the HP theory can be rewritten to arrive to the Driven Liouville-von
Neumann equation plus an additional term (section 2), and then provide a brief
description of the models employed in the simulations (section 3).
In section 4 we examine the impact that the progressive simplifications 
of the HP equation have on the accuracy of the physical description of
these systems.
After that we examine the effect that some of the
relevant parameters of the models have on the baseline
performance of these different approximations.
In particular, we will focus on the $\Gamma$ parameter
(section 5) and the shape of the field used to
generate the reference matrix in the DLvN scheme
(section 6).
Finally, in section 7, we discuss the differences between
the two forms of the DLvN equation reported in the literature:
the one emerging from the truncation of the HP method
(eq. \ref{master-tddft-e}), and the one proposed by Hod and co-workers (eq. \ref{master-tddft-z}).

\section{Recasting the multiple-probe equations of motions}
\label{section.eom}

We consider in this work a system formed by a central molecule or device, 
identified hereafter with the symbol $C$,
coupled to a right and a left electrode (or lead), denoted as regions $R$ 
and $L$ respectively.
In the multiple-probes framework
this system is embedded in an implicitly represented
environment via the coupling of each atom of the left
and right leads to an external probe,
which in isolation has a retarded and an advanced surface Green's function $g^\pm_j (E)$ and a surface
local density of states $d_j(E) = -\pi^{-1} \Im g^+_j(E)$.\cite{hp}
These probes have fixed  electrochemical potentials $\mu_L$ and $\mu_R$, depending on whether they are
connected to the left or right electrode, with Fermi-Dirac distributions $f_L(E)$ and $f_R(E)$.
For this model, and assuming that the Hamiltonian
of the system is time-independent, the HP theory provides
the following equation of motion for the electronic
density (see eq. 38 in reference ~\cite{hp}):
\begin{equation}
i \hbar \dot{\hat{\rho}}_S(t)= [\hat{H}_S, \hat{\rho}_S(t)] + \hat{\Sigma}^+ \hat{\rho}_S(t) 
- \hat{\rho}_S(t) \hat{\Sigma}^-
+ \int [\hat \Sigma^<(E) \hat G^-_S(E) - \hat G^+_S(E) \hat \Sigma^<(E) ] dE
\label{eqhp}
\end{equation}
where $\hat{\rho}$  is the time-dependent density matrix,
$\hat{H}$ is the electron Hamiltonian, subscript $S$ refers
to the full system ($L$, $C$ and $R$ sections), and the
$\hat{\Sigma}$ and $\hat{G}$ matrices are the
system's self-energy and Green's function.
We also assume that (i) all probes interact with the
electrodes through the same coupling term $\gamma$,
and (ii) the wideband limit applies to the probes so that
their density of states becomes a constant, $d_j(E) = d$.
In these conditions, these matrices adopt the
following form:
\begin{equation}
\hat{\Sigma}^\pm = \hat{\Sigma}^\pm_L + \hat{\Sigma}^\pm_R = -\frac{i\Gamma}{2} \cdot \hat P_L - 
\frac{i\Gamma}{2} \cdot \hat P_R = \left( \hat \Sigma^\mp \right)^{\dagger},
\label{sigmapm}
\end{equation}

\begin{equation}
\hat \Sigma^<(E) = \frac{\Gamma}{2\pi} \cdot f_L(E) \cdot \hat P_L +
\frac{\Gamma}{2\pi} \cdot f_R(E) \cdot \hat P_R,
\end{equation}

\begin{equation}
\hat G^\pm_S(E) = \left ( E \cdot \hat P_S - \hat H_S - \hat{\Sigma}^\pm \right )^{-1}.
\label{Gpm}
\end{equation}
In the above equations, $\Gamma = 2\pi \gamma^2 d$,
and $f_L(E)$ and $f_R(E)$ correspond to the
Fermi-Dirac distributions for the left and right probes.
The matrices $\hat P_L$ and $\hat P_R$ are the
corresponding projector operators, with $\hat P_S$
being the projection over the whole explicit system
(i.e. the identity).


We will now take equation \ref{eqhp} as the starting
point for our derivation. Our goal in this section is to express
it in terms of a difference between weighted density
matrices---one corresponding to the current state of
the system and the other to a reference system---that
assumes the form of the driving term in the DLvN
equations.

For this, we focus on the second
($\hat{\Sigma}^+ \hat{\rho}_S(t) - \hat{\rho}_S(t)
\hat{\Sigma}^-$)
and third
($\int [\hat \Sigma^<(E) \hat G^-_S(E) - \hat G^+_S(E)
\hat \Sigma^<(E) ] dE$)
terms on the right hand side of equation \ref{eqhp}, and
examine separately each of the sub-blocks corresponding
to the orbitals of the electrodes ($L$, $R$) and of the
device ($C$).
The first thing to note is that, since all $\hat{\Sigma}$
matrices contain projections on the leads only, the second
and third terms will vanish in the central block:

\begin{equation}
i \hbar \dot{\hat{\rho}}_{CC}(t)= \left ( [\hat{H}_S, \hat{\rho}_S(t)] \right )_{CC}.
\end{equation}

In turn, the off-diagonal blocks of the second term can
be expanded as:

\[
\hat{\Sigma}^+ \hat{\rho}_S - \hat{\rho}_S \hat{\Sigma}^- = -\frac{i \Gamma}{2}
\left ( \hat P_L \hat \rho_S + \hat P_R \hat \rho_S + \hat \rho_S \hat P_L + \hat \rho_S \hat P_R \right ) \]

\begin{equation}
= -\frac{i \Gamma}{2} \left (2 \hat \rho_{LL} + 2 \hat \rho_{LR} + 2 \hat \rho_{RL} + 2 \hat \rho_{RR}
+ \hat \rho_{LC} + \hat \rho_{CL} + \hat \rho_{RC} + \hat \rho_{CR} \right )
\end{equation}
where for conciseness we have omitted the time-dependence
of $\hat \rho$.
In matrix form, this can be written as:

\begin{equation}
\hat{\Sigma}^+ \hat{\rho}_S - \hat{\rho}_S \hat{\Sigma}^- = -\frac{i \Gamma}{2}
\begin{bmatrix}
2 \hat \rho_{LL} & \hat \rho_{LC}  & 2 \hat \rho_{LR} \\
\hat \rho_{CL}  & 0  & \hat \rho_{CR}  \\
2 \hat  \rho_{RL}  & \hat \rho_{RC}  & 2\hat \rho_{RR}  \\
 \end{bmatrix}.
\end{equation}

We now turn our attention to the third term on the right
hand side of equation \ref{eqhp}, considering separately
each off-diagonal block. For the upper left one, we have:

\[
\int [\hat \Sigma^<(E) \hat G^-_S(E) - \hat G^+_S(E) \hat \Sigma^<(E) ]_{LL} dE =
\frac{\Gamma}{2\pi} \int \left(f_L(E) \hat P_L \hat G_S^-(E) \hat P_L - f_L(E) \hat P_L \hat G_S^+(E) \hat P_L\right) dE \]

\begin{equation}
= \frac{\Gamma}{2\pi} \hat P_L \cdot \int f_L(E) \left [ \hat G_S^-(E) - \hat G_S^+(E) \right ] dE \cdot \hat P_L
= i \Gamma \hat P_L \cdot \int f_L(E) \hat D_S(E) dE \cdot \hat P_L
\label{eqLL}
\end{equation}
where we used the relation between the Green's function
and the density of states matrix 
$\hat D_S(E)= (2\pi i)^{-1} \left(\hat G_S^-(E) -
\hat G_S^+(E)\right)$.
The integral in the last term of equation \ref{eqLL}
defines a fictitious equilibrium density matrix with an
electronic distribution $f_L$ corresponding to the left
probes, that we will denote $\hat \rho^{0L}$.
After applying on the left and on the right the
projection operator associated with the upper left
block, we arrive at the final result,

\begin{equation}
\int [\hat \Sigma^<(E) \hat G^-_S(E) - \hat G^+_S(E) \hat \Sigma^<(E) ]_{LL} dE =
i\Gamma \cdot \hat P_L \cdot \hat \rho^{0L} \cdot \hat P_L = i \Gamma \cdot \hat \rho^{0L}_{LL}.
\end{equation}

For the $RR$ sub-matrix we get an analogous expression,
but with the $R$ projectors and the Fermi-Dirac
distribution for the right probes, which defines a
different reference density matrix $\hat \rho^{0R}$,

\begin{equation}
\int [\hat \Sigma^<(E) \hat G^-_S(E) - \hat G^+_S(E) \hat \Sigma^<(E) ]_{RR} dE =
i \Gamma \hat P_R \cdot \int f_R(E) \hat D_S(E) dE \cdot \hat P_R  = i \Gamma \cdot \hat \rho^{0R}_{RR}.
\end{equation}

To evaluate the off-diagonal terms involving the central
region (blocks \textit{LC, CL, RC}, and \textit{CR}),
we first observe that in all these cases part of the
integrand vanishes when operating on $\hat \Sigma^<$
due to the contiguous application of projectors
belonging to different regions,

\[
\int [\hat \Sigma^<(E) \hat G^-_S(E) - \hat G^+_S(E) \hat \Sigma^<(E) ]_{LC} dE =
\int \left( \hat P_L \hat \Sigma^<(E) \hat G_S^-(E) \hat P_C - \hat P_L \hat G_S^+(E) \hat \Sigma^<(E) \hat P_C \right) dE \]

\begin{equation}
= \frac{\Gamma}{2\pi} \int f_L(E) \hat P_L \hat G_S^-(E) \hat P_C dE = 
\frac{\Gamma}{2\pi} \hat P_L \cdot \int f_L(E) \hat G_S^-(E) dE \cdot \hat P_C.
\label{eqLC}
\end{equation}

Similarly,

\begin{equation}
\int [\hat \Sigma^<(E) \hat G^-_S(E) - \hat G^+_S(E) \hat \Sigma^<(E) ]_{CL} dE =
-\frac{\Gamma}{2\pi} \hat P_C \cdot \int f_L(E) \hat G_S^+(E) dE \cdot \hat P_L.
\label{eqCL}
\end{equation}
To recover the density of states from just one of the
Green's matrices, we rewrite the latter in the following
way:
\begin{equation}
\hat G_S^-(E) = \left( \frac{\hat G_S^-(E)}{2} - \frac{\hat G_S^+(E)}{2} \right)
+ \left( \frac{\hat G_S^-(E)}{2} + \frac{\hat G_S^+(E)}{2} \right) = 
i \pi \hat D_S(E) + \hat G_S^\Re (E)
\label{eqg-}
\end{equation}

\begin{equation}
-\hat G_S^+(E) = \left( \frac{\hat G_S^-(E)}{2} - \frac{\hat G_S^+(E)}{2} \right)
+ \left( -\frac{\hat G_S^-(E)}{2} - \frac{\hat G_S^+(E)}{2} \right) =
i \pi \hat D_S(E) - \hat G_S^\Re (E),
\label{eqg+}
\end{equation}
for which we have defined the matrix
$\hat G_S^\Re (E) = \frac{1}{2}(\hat G_S^-(E) +
\hat G_S^+(E))$.
Inserting equations \ref{eqg-} and \ref{eqg+} in
equations \ref{eqLC} and \ref{eqCL}, we arrive at

\[
\int [\hat \Sigma^<(E) \hat G^-_S(E) - \hat G^+_S(E) \hat \Sigma^<(E) ]_{LC} dE =\]

\begin{equation}
\frac{i \Gamma}{2} \hat P_L \cdot \int f_L(E) \hat D_S(E) dE \cdot \hat P_C + 
\frac{\Gamma}{2\pi} \hat P_L \cdot \int f_L(E) \hat G_S^\Re(E) dE \cdot \hat P_C =
\frac{i \Gamma}{2} \hat \rho^{0L}_{LC} + \frac{\Gamma}{2\pi} \hat \Omega^L_{LC}.
\end{equation}
Similarly
\begin{equation}
\int [\hat \Sigma^<(E) \hat G^-_S(E) - \hat G^+_S(E) \hat \Sigma^<(E) ]_{CL} dE =
\frac{i \Gamma}{2} \hat \rho^{0L}_{CL} - \frac{\Gamma}{2\pi} \hat \Omega^L_{CL}
\end{equation}
where we have introduced the matrix
$\hat \Omega^{L/R} = \int f_{L/R}(E) \hat G_S^\Re (E) dE$.
The terms involving the right lead can be treated on
the same footing, to obtain an analogous result but
incorporating the reference density and $\hat \Omega$
corresponding to the right region,

\begin{equation}
\int [\hat \Sigma^<(E) \hat G^-_S(E) - \hat G^+_S(E) \hat \Sigma^<(E) ]_{RC} dE =
\frac{i \Gamma}{2} \hat \rho^{0R}_{RC} + \frac{\Gamma}{2\pi} \hat \Omega^R_{RC}
\end{equation}

\begin{equation}
\int [\hat \Sigma^<(E) \hat G^-_S(E) - \hat G^+_S(E) \hat \Sigma^<(E) ]_{CR} dE =
\frac{i \Gamma}{2} \hat \rho^{0R}_{CR} - \frac{\Gamma}{2\pi} \hat \Omega^R_{CR}.
\end{equation}

To complete the derivation, we consider  the two
remaining blocks that couple both leads together
($LR$, $RL$):

\[
\int [\hat \Sigma^<(E) \hat G^-_S(E) - \hat G^+_S(E) \hat \Sigma^<(E) ]_{LR} dE =
\frac{\Gamma}{2\pi} \int \left(f_L(E) \hat P_L \hat G_S^-(E) \hat P_R - f_R(E) \hat P_L \hat G_S^+(E) \hat P_R\right) dE \]

\begin{equation}
= \frac{\Gamma}{2\pi} \hat P_L \cdot \int \left( f_L(E) \hat G_S^-(E) - f_R(E) \hat G_S^+(E) \right) dE \cdot \hat P_R .
\label{eqLR}
\end{equation}

The expression within the brackets can be rewritten
introducing relations \ref{eqg-} and \ref{eqg+} to
elicit the reference density and omega matrices,
\[
\int \left( f_L(E) \hat G_S^-(E) - f_R(E) \hat G_S^+(E) \right) dE  \]
\[
= \int \left(i\pi f_L(E) \hat D_S(E) + f_L(E) \hat G_S^\Re(E) + i\pi f_R(E) \hat D_S - f_R(E) \hat G_S^\Re \right) dE \]

\begin{equation}
= i\pi \hat \rho_S^{0L} + \hat \Omega_S^L + i\pi \hat \rho_S^{0R} - \hat \Omega_S^R,
\end{equation}
leading to the following result:
\begin{equation}
\int [\hat \Sigma^<(E) \hat G^-_S(E) - \hat G^+_S(E) \hat \Sigma^<(E) ]_{LR} dE =
\frac{i\Gamma}{2} \left( \hat \rho_{LR}^{0L} + \hat \rho_{LR}^{0R} \right) + 
\frac{\Gamma}{2\pi} \left( \hat \Omega_{LR}^L - \hat \Omega_{LR}^R \right)
\end{equation}
which combines the left and right probes populated matrices.
The $RL$ block is treated analogously, to obtain:
\begin{equation}
\int [\hat \Sigma^<(E) \hat G^-_S(E) - \hat G^+_S(E) \hat \Sigma^<(E) ]_{RL} dE =
\frac{i\Gamma}{2} \left( \hat \rho_{RL}^{0L} + \hat \rho_{RL}^{0R} \right) + 
\frac{\Gamma}{2\pi} \left( \hat \Omega_{RL}^R - \hat \Omega_{RL}^L \right).
\end{equation}
Thus,  the third term in equation \ref{eqhp} can in
matrix form be written as:
\[
\int [ \hat \Sigma^<(E) \hat G^-_S(E) - \hat G^+_S(E) \hat \Sigma^<(E) ] dE \]

\begin{equation}
= \frac{i\Gamma}{2}
\begin{bmatrix}
2\hat \rho^{0L}_{LL}  & \hat \rho^{0L}_{LC}  & \hat \rho^{0L}_{LR} + \hat \rho^{0R}_{LR} \\
\hat \rho^{0L}_{CL}   &  0  & \hat \rho^{0R}_{CR}  \\
\hat \rho^{0L}_{RL} + \hat \rho^{0R}_{RL} & \hat \rho^{0R}_{RC}  & 2\hat \rho^{0R}_{RR}  \\
\end{bmatrix}
+\frac{\Gamma}{2\pi}
\begin{bmatrix}
0 & \hat \Omega^L_{LC} & \hat \Omega^L_{LR} - \hat \Omega^R_{LR} \\
-\hat \Omega^L_{CL} & 0 & -\hat \Omega^R_{CR} \\
\hat \Omega^R_{RL} - \hat \Omega^L_{RL} & \hat \Omega^R_{RC} & 0 \\
\end{bmatrix}
\end{equation}
where the first matrix on the right hand side will be
referred to as the ``region-weighted reference density
matrix'', or simply reference density matrix.

Finally, collecting all the pieces together, the HP
master equation can be rewritten in the following way:
\[
 \dot{\hat{\rho}}(t) = -\frac{i}{\hbar} [\hat{H}_S, \hat{\rho}_S(t)] 
- \frac{\Gamma}{2\hbar}
\begin{bmatrix}
2(\hat \rho_{LL} - \hat \rho^{0L}_{LL}) & \hat \rho_{LC} - \hat \rho^{0L}_{LC} & 2\left(\hat \rho_{LR} - 
\frac{\hat\rho^{0L}_{LR} + \hat \rho^{0R}_{LR}}{2}\right)\\
\hat \rho_{CL} - \hat \rho_{CL}^{0L} & 0 & \hat \rho_{CR} - \hat \rho_{CR}^{0R} \\
2\left(\hat \rho_{RL} - \frac{\hat\rho^{0L}_{RL} + \hat \rho^{0R}_{RL}}{2}\right) & \hat \rho_{RC} - \hat \rho^{0R}_{RC} &
2(\hat \rho_{RR} - \hat \rho^{0R}_{RR}) \\ 
\end{bmatrix} \]

\begin{equation}
- \frac{i\Gamma}{2\pi \hbar}
\begin{bmatrix}
0 & \hat \Omega_{LC}^L &  \hat \Omega_{LR}^L - \hat \Omega_{LR}^R \\
-\hat \Omega^L_{CL} & 0 & -\hat \Omega^R_{CR} \\
\hat \Omega^R_{RL} - \hat \Omega^L_{RL} & \hat \Omega^R_{RC} & 0 \\
\end{bmatrix}
\label{hp-rewritten}
\end{equation}

We have thus arrived at a mathematical structure that
is very similar to that of the DLvN scheme as presented
in equation \ref{master-tddft-e}.
By this we mean that the second term has the form of a
block-weighted difference between the current time
dependent density matrix and a reference density matrix.
The implications of the definition of this
reference matrix and the impact of including or excluding
the remaining $\Omega$-term are analyzed in the remainder of
this work.


\section{Equations of motion, model systems, and time propagation}
\label{section.model}

In what follows we assess and compare the performance of three
different implementations of the HP equation of motion:
(i) the full formula \ref{hp-rewritten};
(ii) a version excluding the $\Omega$-term; and
(iii) same as (ii), but where the region-weighted
reference density matrix is computed using a step-shaped
field encoding the bias potential.
In the latter, notice that $\hat \rho^{0L}$ and
$\hat \rho^{0R}$ are the same matrix $\hat \rho^0$,
corresponding to the polarized density in the presence
of a step-field, and thus the equation of motion becomes
identical to equation \ref{master-tddft-e}.
These three schemes (i), (ii) and (iii), will be
referred to as ``full-HP'', ``partial-HP'', and
``step-potential'' methods, or, for short, F-HP,
P-HP, and ST-P, respectively.
As mentioned above, the mathematical structure of the
P-HP approach resembles closely the one of the DLvN
equation \ref{master-tddft-e}, implemented in this
study as the ST-P method.
In particular, the driving term arising in the HP
formula reinstates the damping of the coherences
to the equilibrium values.
In this sense, the ST-P scheme can also be thought
of as an approximation to the P-HP method with an
alternative reference density matrix.
The motivation to explore it is to bypass the
calculation of Green's functions, which becomes
especially appealing in the context of ab-initio
Hamiltonians.

For comparison purposes, we also consider in this work
the alternative form of DLvN given by equation
\ref{master-tddft-z}, which hereafter will be referred to as DLvN-z, whereas
the version in equation \ref{master-tddft-e} will be denoted as DLvN-e
(since these two equations damp the lead-molecule coherences to zero and to their equilibrium
values, respectively).
It is important to note from this is that ``ST-P''
and ``DLvN-e'' refer to the exact same method:
we will in general use the first notation when comparing only with
other HP-derived methods, and the second one when also
including the other version of the DLvN equation.

To examine the performance of these different equations of motion, a
series of model systems were chosen as case studies in numerical simulations.
All these systems consist of linear chains of atoms
arranged in different spatial configurations (see below).
The electronic structure is represented
by a tight-binding Hamiltonian with an
orthonormal one-electron basis, using
the  model presented by Sutton and co-workers
\cite{sutton_tb} (other possible, more elaborate
electronic models are discussed in~ \cite{ref25-horsfield2016efficient}).
This scheme adopts a single orbital $|j\rangle$ per
atom (or site), coupled with each other
by  nearest-neighbour hopping terms that only
depend on the distance between sites.
The onsite energies will in general be identical for
all atoms, except in 
the ``disordered'' model (see below), or in the presence
of an external field used to obtain the reference matrix.

For the HP method, the potential difference
$\Delta V$ between the left and right probes determines the
Fermi-Dirac distributions according to the corresponding
electrochemical potentials $\mu_L = + \Delta V / 2$
and $\mu_R = - \Delta V / 2$.
The presence of an external field (to generate the reference density
to be used either in the modified HP or the DLvN schemes) is simulated through
the modification of the onsite energies.
The shape of this field is a matter of analysis in
section 6, but for the simulations in sections 4 and 5, constant
values of $+ \Delta V / 2$ and $- \Delta V / 2$
were used for the atoms in the left and right leads
(respectively), whereas the onsite energies of the
atoms in the device were left unmodified.

The time integration of the equations of motion for the
electrons was performed by using a simple Leapfrog
algorithm, and for all the simulations presented in
this work the time-step chosen was of 0.005 fs.
The driving term was weighted by an extra time dependent
factor that grows linearly from 0 to 1 in
the first few steps of the propagation.
Tests were performed to check that the time step and the
slope of the initial ramp did not affect the shape
of the current.

\begin{figure} 
\begin{center}
\includegraphics[scale=0.75,keepaspectratio=true]{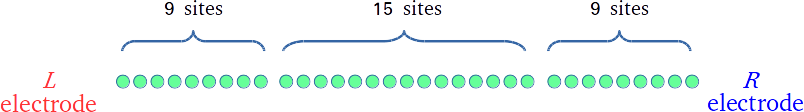}
\end{center}
\caption{Scheme of the resonant system.
All on-site energies are the same.}
\label{resonant-scheme}
\end{figure}

Simulations were performed on three basic systems
characterized by the configuration of their central
region ($C$).
In all systems the electrodes $L$ and $R$
are made of 200 atoms, except for some results shown
in section 7 in which shorter leads 
of 50 atoms were explored.
In all cases the systems are metallic,
exhibiting a half-filled conduction band.
These three systems are:

\begin{itemize}
\item Ballistic: all atoms in the system ($S$)
are equally spaced by 2.5 $\AA$, involving a
hopping integral of -3.88 eV (which corresponds
to the tight-binding reference parameters for
gold\cite{sutton_tb}).
The on-site energy is the same for all atoms.
There are 21 atoms in the central region $C$.

\item Disordered: this system has the exact same
geometry as the ballistic (same number of atoms
all equally spaced), but with a different on-site
energy for every atom belonging to the central
region $C$.
These different on-site energies were randomly
generated using a homogeneous distribution with
values between plus and minus the absolute value
of the hopping term above.

\item Resonant: all sites are equally spaced as
in the previous cases, with the exception of two
distances  separating a group of 15 atoms in the
center of the device, from two groups of 9 atoms
on each end, connected to the electrodes.
All of these belong to the $C$ region, as can be
seen in Figure \ref{resonant-scheme}. 
These two bonds were 0.5 \AA~longer than the rest.
The on-site energies are the same for all atoms,
but the hopping terms of the longer bonds are
-1.88 eV.
\end{itemize}

Modified versions of the resonant system were later
used in section 6.
Instead of having a 9-15-9 configuration of the central
region (with the dash standing for the longer separation),
these are 9-45-9 and 30-15-30.

\section{The components of the driving term and their role}
\label{section.perform}

\begin{figure}
\begin{center}
\includegraphics[scale=0.35,keepaspectratio=true]
{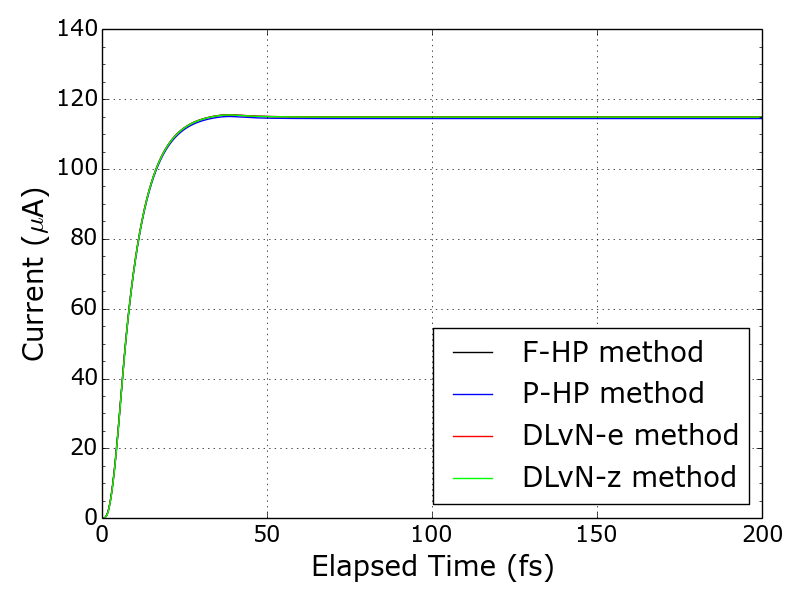}
\includegraphics[scale=0.35,keepaspectratio=true]
{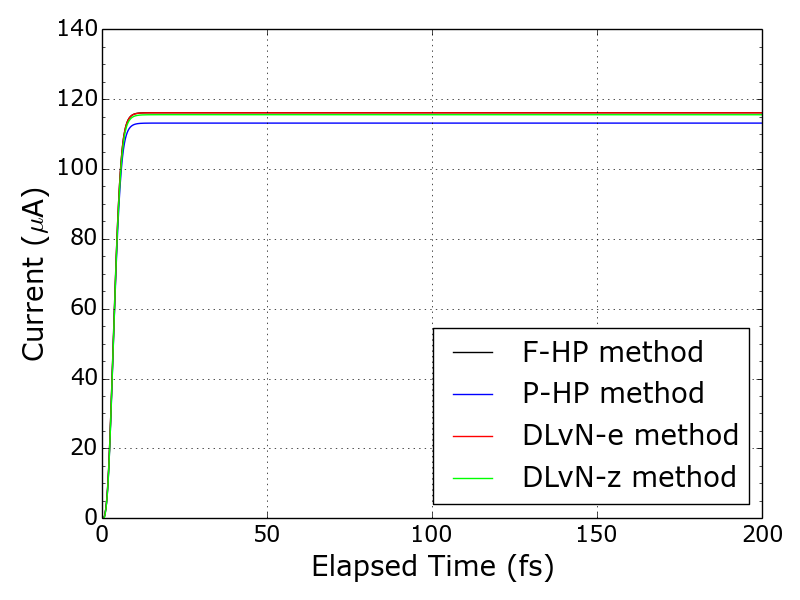}\\
\includegraphics[scale=0.35,keepaspectratio=true]
{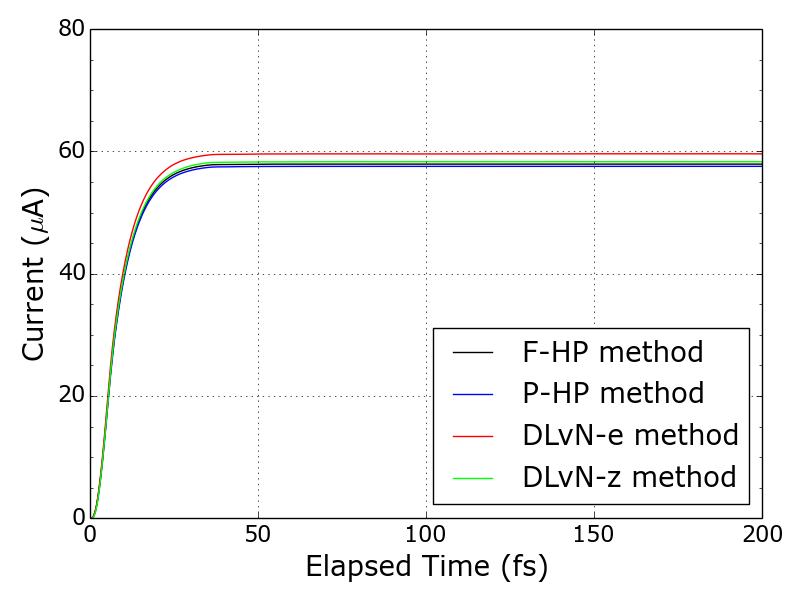}
\includegraphics[scale=0.35,keepaspectratio=true]
{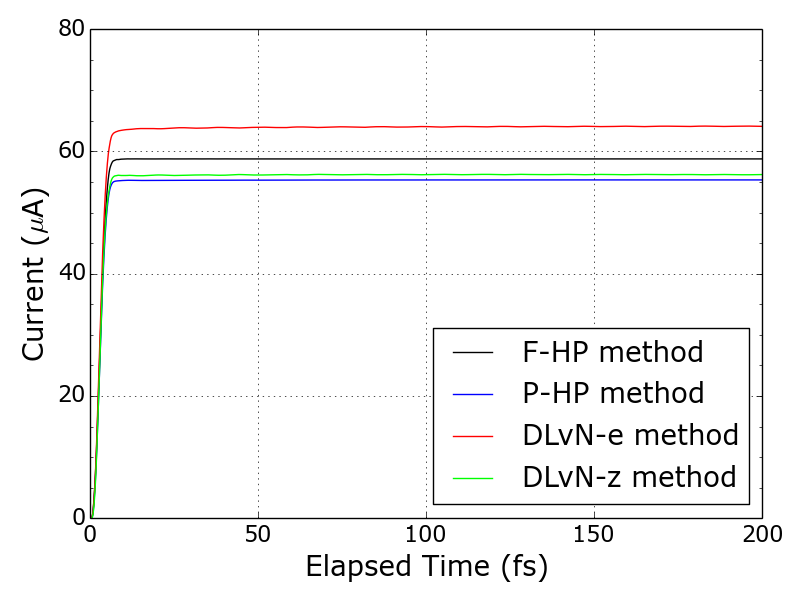}\\
\includegraphics[scale=0.35,keepaspectratio=true]
{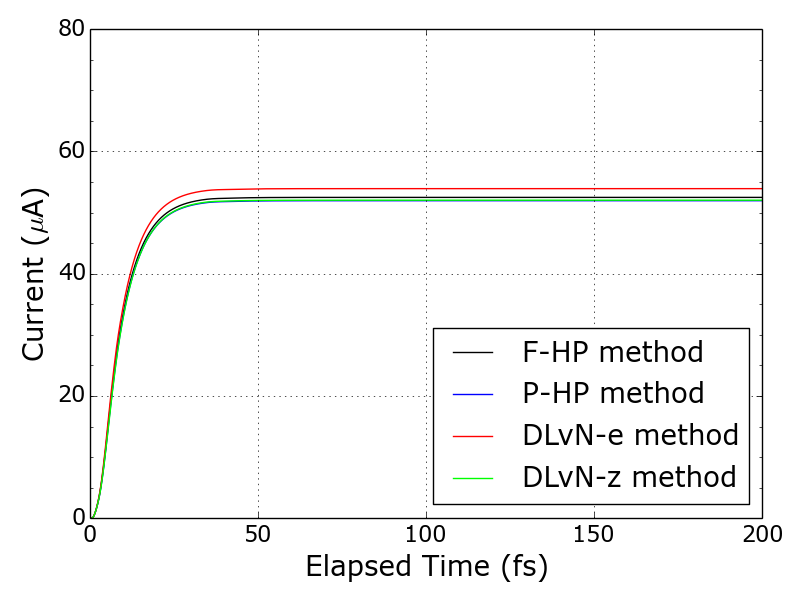}
\includegraphics[scale=0.35,keepaspectratio=true]
{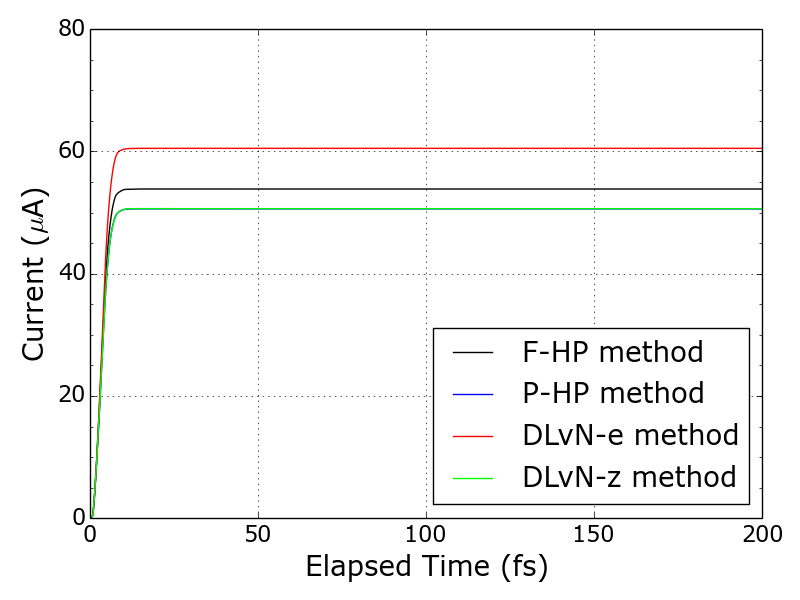}
\end{center}

\caption{Current as a function of time obtained for
a bias of 1.5 V with the three different implementations
of the hairy-probes method (including the so called
DLvN-e, given by equation \ref{master-tddft-e}), and
with the Driven Liouville-von Neumann formula given
in equation \ref{master-tddft-z} (DLvN-z).
The upper, middle and lower panels, correspond to the
ballistic, the  disordered, and the resonant models,
respectively.
Left and right panels show the currents for $\Gamma$
equal to 0.1 eV and 0.6 eV.}
\label{STAB}
\end{figure}

Figure \ref{STAB} depicts the current as a function of time for transport simulations based
on the F-HP, P-HP, ST-P (or DLvN-e), and DLvN-z methods.
It is seen that the currents reach a stable steady
state, regardless of the method, model system, and the
value of $\Gamma$.
The upper, middle and lower panels of Figure \ref{STAB},
correspond to the ballistic, the resonant, and the
disordered models, respectively, for an applied bias of
1.5 V (the same stable behavior is found for other bias).
Left and right panels show the currents for $\Gamma$
equal to 0.1 eV and 0.6 eV.
It can be seen that for large couplings, the steady
state is reached faster, and that in none of the cases
is there any ambiguity in the identification of the
final current, since it typically converges to a precise
value within the first 10 - 40 fs of dynamics, depending
on $\Gamma$.
While the figures present the current for the initial
400 fs, most of the simulations have been evolved for
up to 1 ps, without the detection of instabilities.

\begin{figure}
\begin{center}
\includegraphics[scale=0.42,keepaspectratio=true]{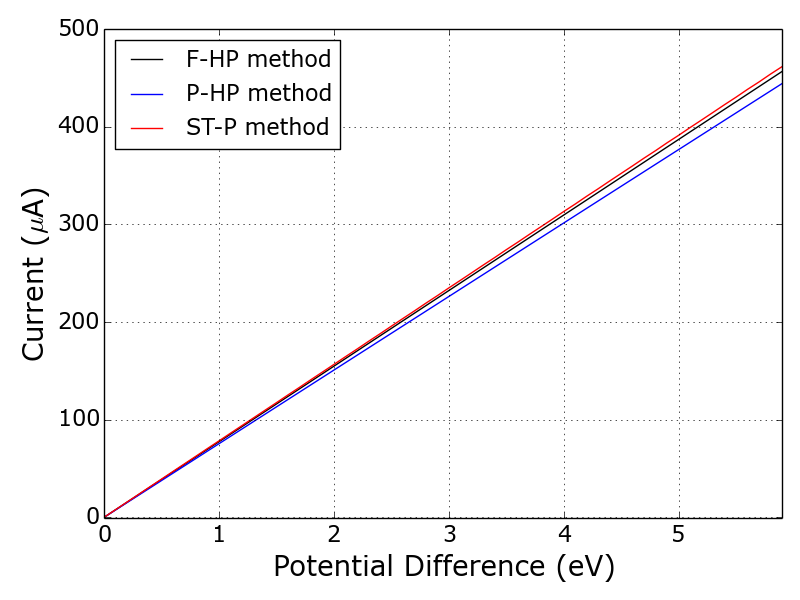}\\
\includegraphics[scale=0.42,keepaspectratio=true]{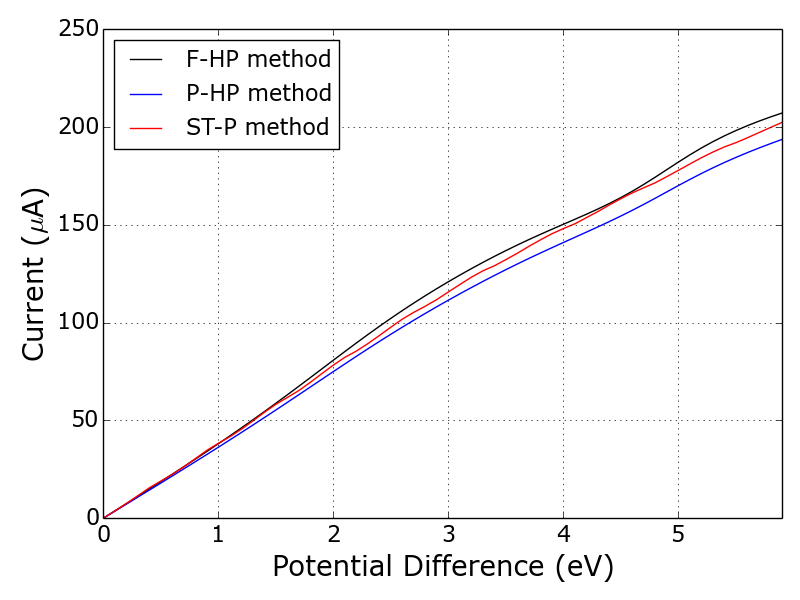}\\
\includegraphics[scale=0.42,keepaspectratio=true]{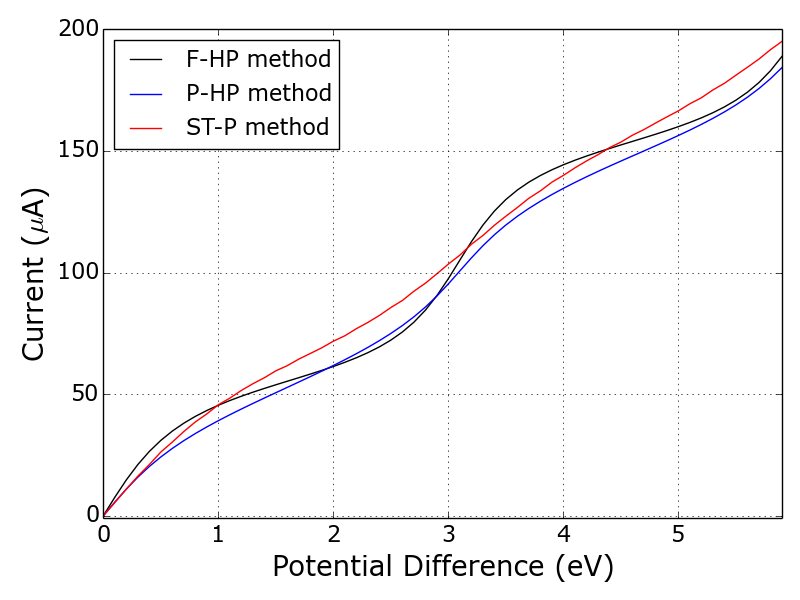}\\
\end{center}
\caption{Current as a function of the applied bias,
computed in the steady state from quantum dynamics
simulations based on the three different implementations
of the hairy-probes method.
Results are shown for the ballistic (top), disordered
(center), and resonant (bottom) model systems.}
\label{IvsV}
\end{figure}

Figure \ref{IvsV} presents the steady state
current-voltage curves obtained from the three
methods deriving from HP.
For the ballistic system all three approaches show
only marginal differences between each other, which
become negligible in the low bias region.
The same is true for the disordered case.
In both kinds of model system, the differences in the
steady-state currents resulting from the three methods
are below 5\%, and typically much less than this at low
biases.
Somehow curiously, the scheme based on the step-potential
reproduces more accurately the response of the full HP
method, despite the difference in their reference densities.
This suggests the existence of a compensation mechanism
through which the step-generated reference density matrix
balances the absence of the $\hat \Omega$ contribution.
A possible explanation can be found in its higher
polarization, as discussed below.

\begin{figure}
\begin{center}
\includegraphics[scale=0.3,keepaspectratio=true]
{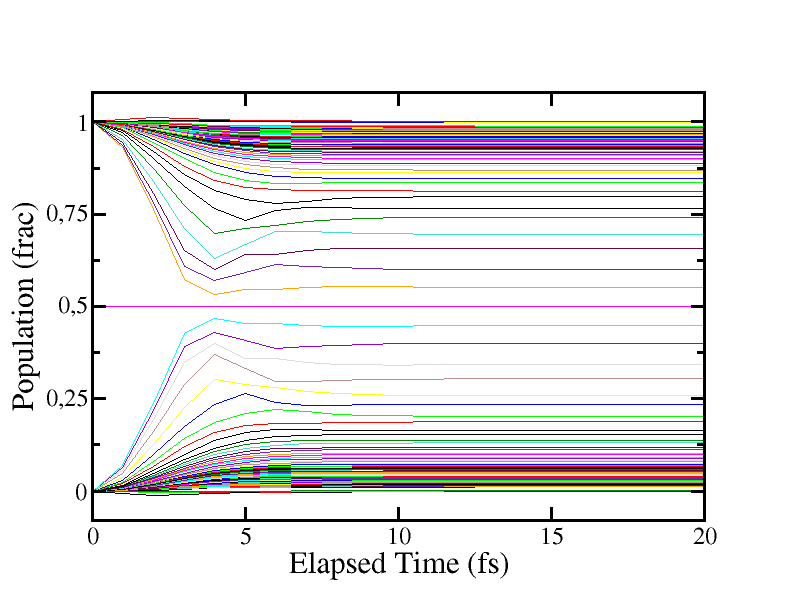}\\
\includegraphics[scale=0.3,keepaspectratio=true]
{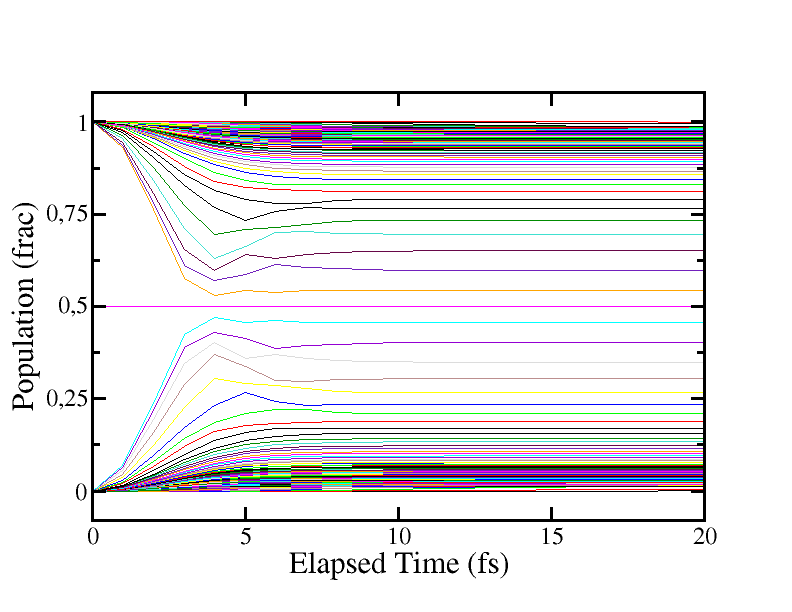}\\
\includegraphics[scale=0.3,keepaspectratio=true]
{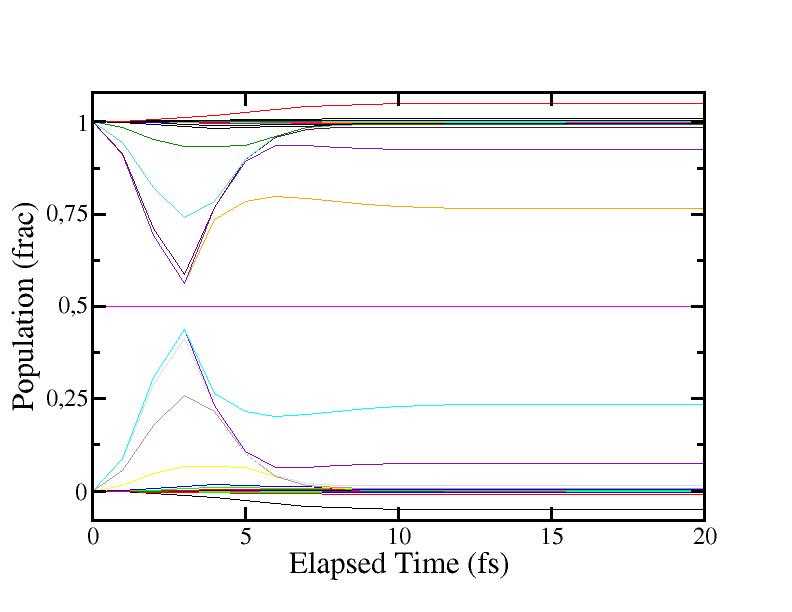}\\
\end{center}
\caption{Temporal evolution of the eigenvalues of the
density matrix for the resonant model system, computed
using the full HP (top), P-HP (center), and ST-P
(bottom) methods.}
\label{eigenstatespop}
\end{figure}

The $I-V$ curves in the resonant system display a more
complex behavior. The P-HP and ST-P equations of motion
still reproduce the currents of the full HP method, but
with slightly larger deviations.
Here, these appear to be comparable in the P-HP and in
the ST-P method. The latter scheme produces in all three
systems, for a given bias, higher currents than the first
one.
This trend was confirmed also in exploratory calculations
for other resonant systems with variable separation
between atoms: the step-potential yields systematically
larger steady states currents than the P-HP method for
a fixed potential difference.
This might be related to the fact that in the case of a
reference density generated from an abrupt step-field,
the device is subject to a larger effective bias than
in the case of a smoother reference density constructed
from the Green's functions.
In other words, the action of the field, through the
direct modification of the on-site energies, affects
more severely the density matrix of the electrodes when
compared to the HP method, in which the electrochemical
potential at the leads is controlled via the coupling
with external probes according to $\gamma$.
This  is consistent with the eigenstates populations
plotted in Figure \ref{eigenstatespop}.
The full and the P-HP simulations produce a manifold of
partially occupied states, both methods exhibiting
essentially no differences from each other.
When the reference density derives from the step-wise
potential, instead, the eigenvalues of the density
matrix are  0 or 1, with a small fraction of states
associated with intermediate occupations.
This circumstance  reflects a lower electronic
temperature and a more drastic polarization in the
ST-P approach arising from a highly perturbed
$\hat \rho^0$, in comparison with the other two
treatments.
On the other hand, in the former there are a few states
that violate the Pauli's principle, bearing negative,
or larger than unity occupations.
The results in Figure \ref{eigenstatespop} correspond
to  the resonant system, but  analogous behavior is
found for the ballistic and resonant models.
The behavior of the eigenvalues for the different schemes,
including the one in equation \ref{master-tddft-z}, is
further discussed in the final section.

\section{Effect of the $\Gamma$ parameter}
\label{section.gamma}

The question of the proper value of $\Gamma$ is
an interesting problem that has been investigated
recently,\cite{ref25-horsfield2016efficient}
but has not been entirely settled.
In the context of the DLvN approach, it has been argued
in the literature that the magnitude of the driving rate
should be somewhere between  the energy spacing separating
the lead levels, and the effective device-lead coupling.
\cite{franco}
If  $\Gamma$ is too high, the target density is enforced
in the lead sections too tightly to allow for an
appropriate response and relaxation at the interface.
As a consequence, in this regime the current is a
decreasing function of $\Gamma$ (see e.g. refs. ~\cite{franco} , ~\cite{jctc_10_2927} , ~\cite{jcp_morzan}).
Moreover, when $\Gamma$ is too low,  the relaxation
process in the leads is not sufficiently fast to allow
for the electronic density in the leads to reach
$\hat \rho^0$.
The net effect in this case can be assimilated to a
situation in which the leads are disconnected or weakly
connected to the reservoirs and the system approaches a
microcanonical regime.

On the other hand, in the context of hairy probes
$\Gamma$ depends on the system parameters $d$ and
$\gamma$, where $d$ is the probe surface density of
states and $\gamma$ the matrix element coupling the
sites at the electrodes with the external probes.
While the value of $d$ can be defined to fit the
density of states of the device, there is no particular
criterion to uniquely assign the value of $\gamma$.
In principle, in the wideband limit it is expected to
satisfy the second requirement listed above for $\Gamma$
in DLvN: it must be larger than the mean energy level
spacing in the leads, to ensure an effective broadening
of the electronic states and hence a metallic behavior
of the electrodes.
In the present case this implies a lower bound for
$\Gamma$ of around 0.1 eV.

\begin{figure}
\begin{center}
\includegraphics[scale=0.4,keepaspectratio=true]{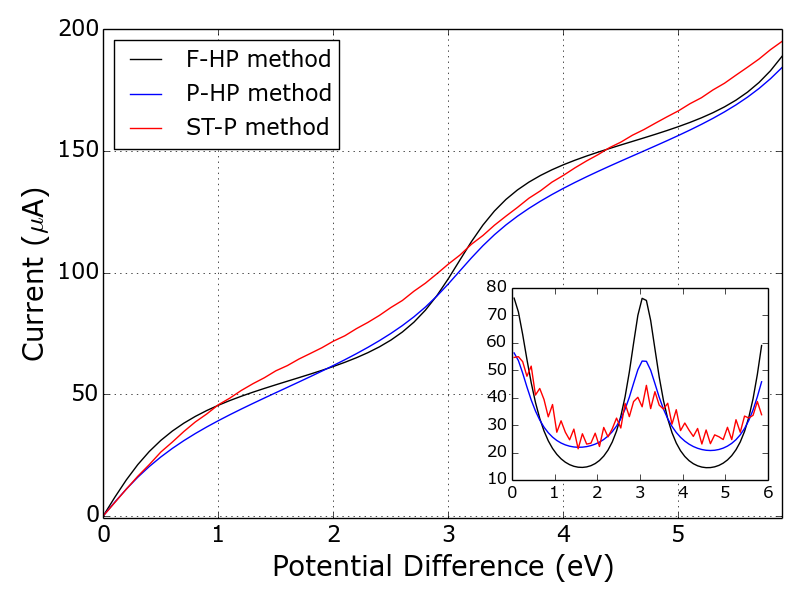}
\includegraphics[scale=0.4,keepaspectratio=true]{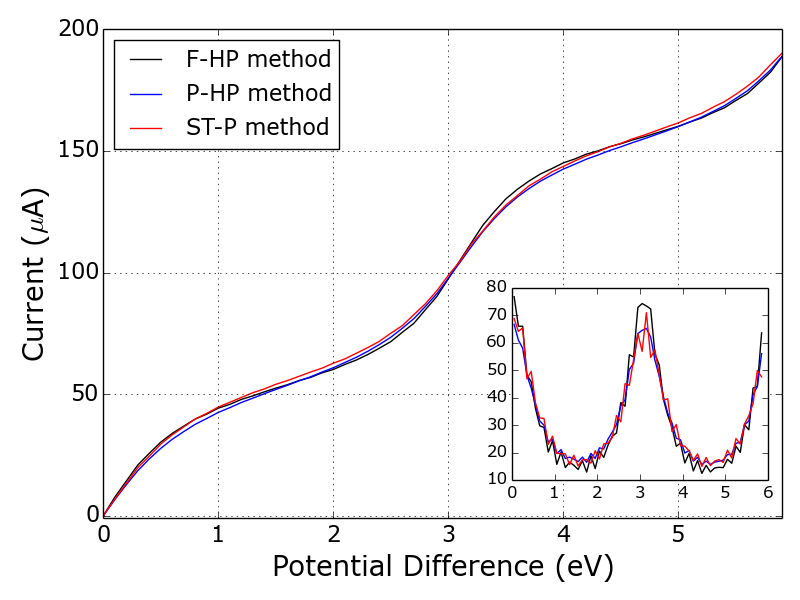}
\end{center}
\caption{Steady state current as a function of
the applied bias computed with the three different
implementations of the hairy-probes method, for
two different values of the $\Gamma$ parameter:
0.6 eV (left panel) and 0.1 eV (right panel).
The results correspond to the resonant system.
For a better comparison, the derivatives are
shown in the insets.}
\label{Gamma}
\end{figure}

Figure \ref{Gamma} illustrates the effect of $\Gamma$
on the $I-V$ plots in the case of the resonant system,
also showing the derivatives of the current (conductance)
in the insets.
We chose this system because in its case the $I-V$ curve displays a distinctive
physical pattern, and the discrepancies between the three approaches
are more significant than in any other model, both
facts making improvements  easier
to identify.
Indeed, these discrepancies tend to fade away as $\Gamma$
decreases from 0.6 to 0.1 eV (within this range $\Gamma$
remains larger than the mean energy level spacing in the
leads, of 0.078 eV).
Interestingly, the insets reveal that conductances
derived from the ST-P methodology turn out to be more
rugged  than in the other methods. 
This can be attributed to the presence of the self-energy
in the Green's functions, that, provided $\Gamma$ is
larger than the energy level spacing in the electrodes,
screens the far ends of the system in the case of the
F-HP and P-HP methods.
In the absence of  the self-energy, finite size effects
manifest in the form of interference oscillations.
As a matter of fact, for a small $\Gamma$ the wiggles
are visible in all three curves (see the inset on the
right panel of Figure \ref{Gamma}), because the screening
effect dilutes as the Green's function carries a
dependence on this parameter in the denominator.
The insets also reveal that a large $\Gamma$ value
tends to damp the resonances in the ST-P case, which
are otherwise intact for $\Gamma$ = 0.1 eV.

\begin{figure}
\begin{center}
\includegraphics[scale=0.46,keepaspectratio=true]{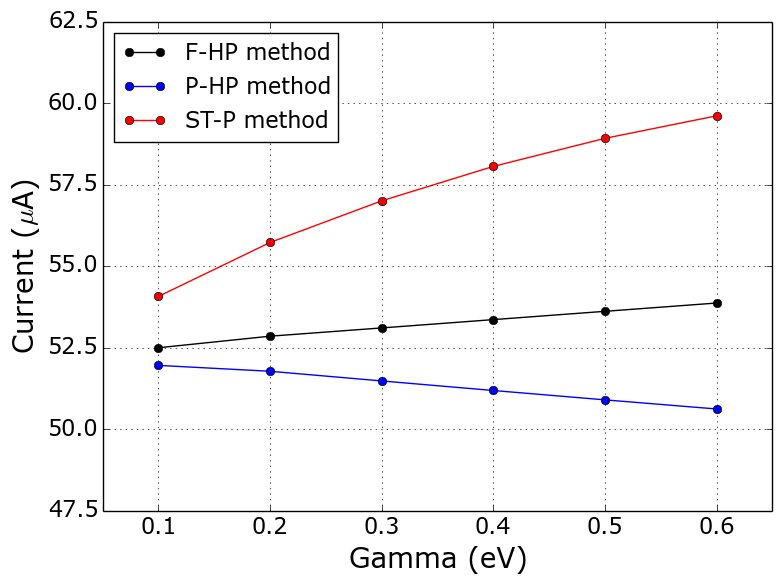}\\
\end{center}
\caption{Steady state current as a function of the
$\Gamma$ parameter, computed with the three different
implementations of the hairy-probes method.
The shown results correspond to the resonant system
with an applied bias of 1.5 V.}
\label{Gamma-1.5}
\end{figure}

The currents obtained for a bias of 1.5 V are depicted
as a function of $\Gamma$ in Figure \ref{Gamma-1.5},
where it can be seen that the agreement between methods
is progressive as the parameter gets smaller.
Additionally, this Figure shows that all approaches
are relatively robust against the variation of
$\Gamma$, in particular the full HP scheme, for which
the change in the steady state current is just 2\% when
this parameter is  reduced by a factor of six. 

\begin{figure}
\begin{center}
\includegraphics[scale=0.38,keepaspectratio=true]{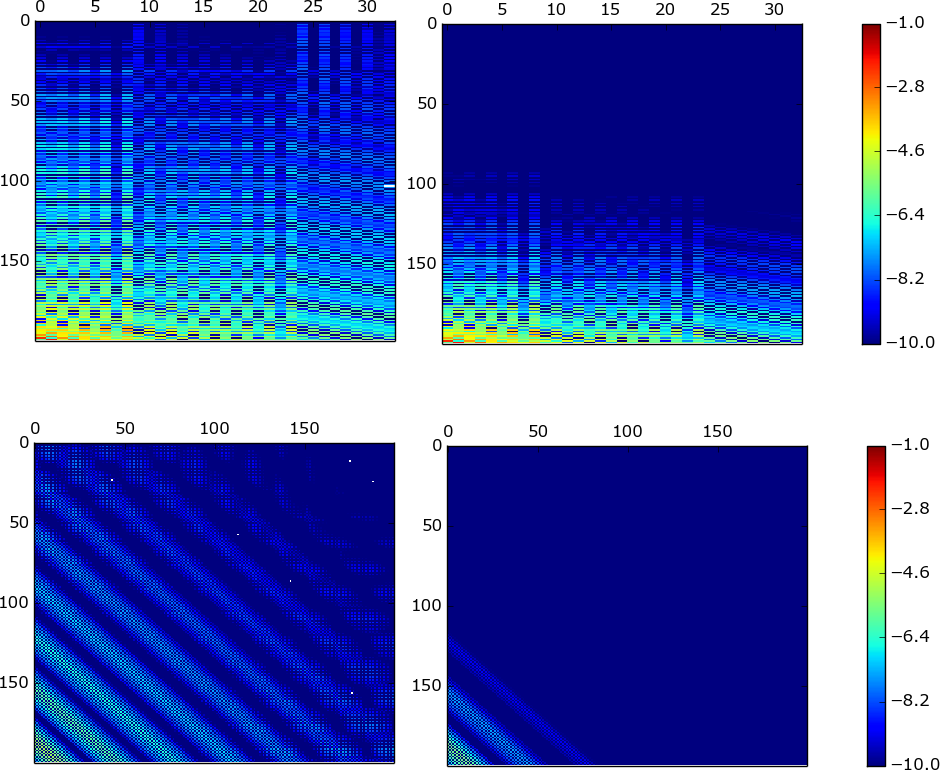}\\
\end{center}
\caption{Color maps representing the structure of
different blocks of the $\hat \Omega$ matrix for
different values of the $\Gamma$ parameter, computed
for the resonant system.
Top panels: LC blocks. Bottom panels: LR blocks.
Left panels: $\Gamma$=0.1 eV. Right panels:
$\Gamma$=0.6 eV.
The colors reflect the absolute values of the matrix
elements, in a natural logarithmic scale.
This matrix determines the difference between the F-HP
and P-HP schemes.}
\label{matrices1}
\end{figure}

\begin{figure} 
\begin{center}
\includegraphics[scale=0.55,keepaspectratio=true]{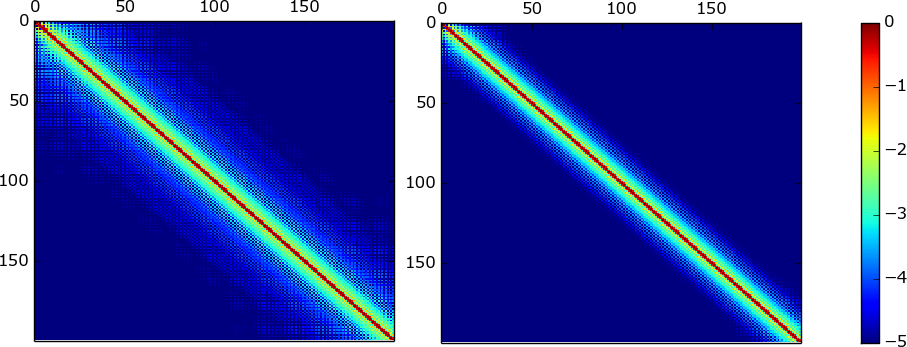}\\
\end{center}
\caption{Color maps representing the LL block of the
reference density matrix $\hat \rho^0$ in the F-HP and
P-HP methods, for different values of the $\Gamma$
parameter, in the case of the resonant system.
Left panel: $\Gamma$=0.1 eV. Right panel: $\Gamma$=0.6 eV.
The colors reflect the absolute values of the matrix
elements, in a natural logarithmic scale.}
\label{matrices2}
\end{figure}

In the F-HP and P-HP dynamics, the effect of $\Gamma$
arises not only from the multiplication of the driving
terms, but also because the reference matrices
$\hat \rho^0$ and $\hat \Omega$ are all functions of
$\Gamma$ through the self-energy and the Green's function
(see equations \ref{sigmapm} and \ref{Gpm}).
Therefore, it becomes relevant to check how the
structure of these matrices change when varying the
value of $\Gamma$.
Figures \ref{matrices1} and \ref{matrices2} show that
the elements of these matrices exhibit a slight inverse
dependence on $\Gamma$, which, as mentioned above, limits
the spatial range of the Green's functions and reference
density matrices in the electrode region, in a way that
has no analogue in ST-P.
This spatial damping is visible on Figures \ref{matrices1}
and \ref{matrices2}.

In particular, the magnitude of the $\hat \Omega$
contribution is responsible for the difference between
the F-HP and P-HP methods. 
All diagonal blocks of the $\hat \Omega$ matrix are
identically zero, and hence only the LC and LR blocks
are depicted in Figure \ref{matrices1} (the remaining
blocks, CL, RL, RC and CR are similar).
The off-diagonal elements of $\hat \Omega$ are of the
same order of magnitude as those of the reference matrix
$\hat \rho^0$. Yet, the impact of the former matrix on
the dynamics is only secondary because it has zeros
throughout its diagonal blocks, which is where
$\hat \rho^0$ gives the highest contribution.
In spite of its seemingly small importance in numerical
terms, though, the presence of the $\hat \Omega$ matrix
appears to balance the effect of $\hat \rho^0$ in the
current, furnishing the full HP method with a reduced
sensitivity with respect to $\Gamma$.

\begin{figure}
\begin{center}
\includegraphics[scale=0.55,keepaspectratio=true]{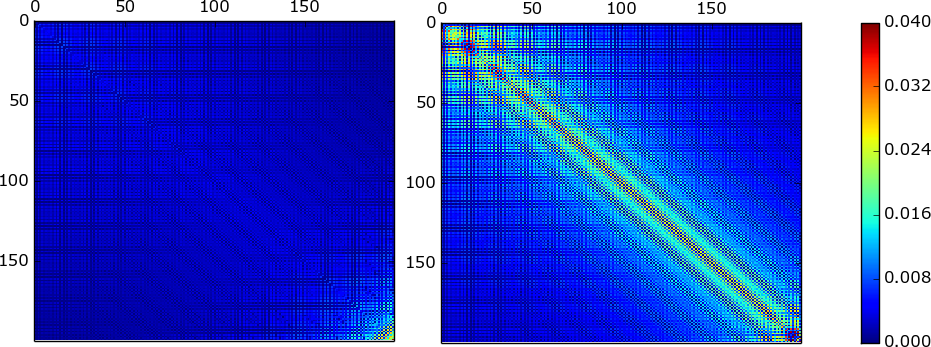}\\
\end{center}
\caption{Color maps representing the LL block for the
difference between the reference density matrices of
the P-HP  and ST-P methods, for the resonant system
and different values of the $\Gamma$ parameter.
Left panel: $\Gamma$=0.1 eV. Right panel: $\Gamma$=0.6 eV.
The colors reflect the absolute values of the matrix
elements.}
\label{PvsS}
\end{figure}

Figure \ref{PvsS} displays the difference between the
reference density matrices in the P-HP and ST-P methods.
Given that in the latter this matrix is independent of
$\Gamma$, Figure \ref{PvsS} just reflects the effect of
this parameter on the hairy probes reference density.
Interestingly, since the decrease of $\Gamma$ tends to
relax the spatial damping of the P-HP reference density,
which is in any case absent from the step-field generated
$\hat \rho^0$, both matrices  become with such a decrease
more similar to each other.
The agreement between these two approaches at low
$\Gamma$ is manifest in the behavior of the currents
in Figures \ref{Gamma} and \ref{Gamma-1.5}. 

As mentioned in the introduction, Zelovich and co-workers 
proposed a method for the computation of a set of broadening factors that
are applied to the lead states. In this way, the rate parameter $\Gamma$ is replaced by diagonal matrices
$\hat \Gamma_L$ and $\hat \Gamma_R$, with dimensions given by the number of basis sets associated with the
leads. The calculation of these matrices involves a  self-consistent procedure to extract from the self-energy
of the isolated lead, the broadening factors that afterwards must be assigned to the corresponding
levels of the lead coupled to the reservoir. Whereas this procedure is somehow involved,
the $\hat \Gamma$ matrices are transferable to any calculation using the same lead model,
and the propagation of the dynamics represents a negligible additional computational cost.
The authors rationalized the magnitude of the resulting broadening factors $\Gamma_i$ by invoking Fermi's
golden rule for a single lead level coupled to a reservoir, which gives a value determined essentially
by the coupling matrix element between  lead and reservoir states. This is precisely the meaning of $\gamma$
in the context of hairy probes.
In any case, this treatment was shown to be of importance in particular
when the DOS of the leads is inhomogeneous in the vicinity of the Fermi energy. However,
the effect of considering a single rate parameter instead of one per level was shown  negligible in
simple tight-binding models as the present one, where the DOS is uniform around the Fermi energy.
As a matter of fact, the authors reported that the adoption of the maximal broadening value calculated for such systems
using their procedure  as the single driving
rate, produces  current traces and steady-state occupations
almost indistinguishable from those obtained using the parameter-free DLvN method.

\section{Generation of the reference density}
\label{section.reff}

The encouraging results obtained with the ST-P
implementation raise the question of whether it
is possible to better reproduce the F-HP dynamics
with a truncated equation of motion, through the
optimization of the reference density.
This pathway is considered in the present
section, using the shape of the electric field as
a tool to prepare $\hat \rho^0$, although other
strategies could be adopted as for example the
state representation of Hod and co-workers.
\cite{jctc_10_2927}

\begin{figure}
\begin{center}
\includegraphics[scale=1.1,keepaspectratio=true]{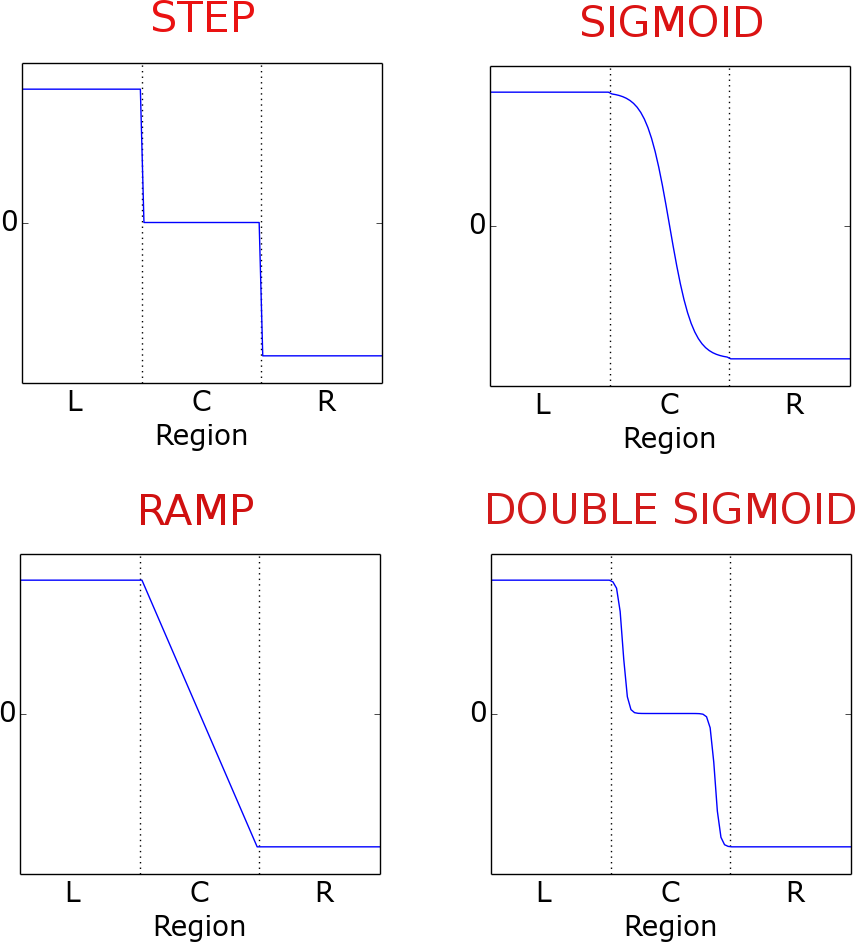}\\
\end{center}
\caption{Shapes of the different electric fields
applied to generate the reference density. L, C,
and R, represent the left electrode, central, and
right electrode regions, respectively.}
\label{field-patterns}
\end{figure}

\begin{figure}
\begin{center}
\includegraphics[scale=0.55,keepaspectratio=true]{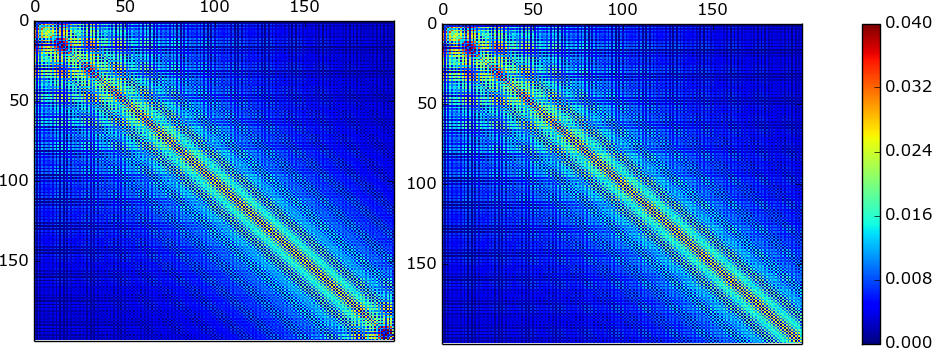}\\
\end{center}
\caption{Color maps representing the LL block
for the  difference between the reference density
matrix of the full HP method with the one generated
with a step field (left), or the one generated with
a sigmoidal field (right).
Data corresponding to the resonant system for
$\Gamma$=0.6 eV. The colors reflect the absolute
values of the matrix elements.}
\label{map-sigmoid}
\end{figure}

We initially compare the results obtained with a
step-like potential, with those corresponding
to  a ramp, a sigmoidal decay, and a double sigmoidal
decay at the start and the end of the central region.
These different patterns are illustrated in Figure
\ref{field-patterns}.
The analysis of the reference matrices generated in
this way did not reveal any significant improvement
in the description of the full hairy probes density,
with the exception of a very slight increase in
similarity for those matrix elements lying in the
proximity of the central region.
This improvement was most noticeable in the case of
the sigmoid field (denoted as ``SI-P''), as can be
visualized in Figure \ref{map-sigmoid}.

\begin{figure}
\begin{center}
\includegraphics[scale=0.42,keepaspectratio=true]{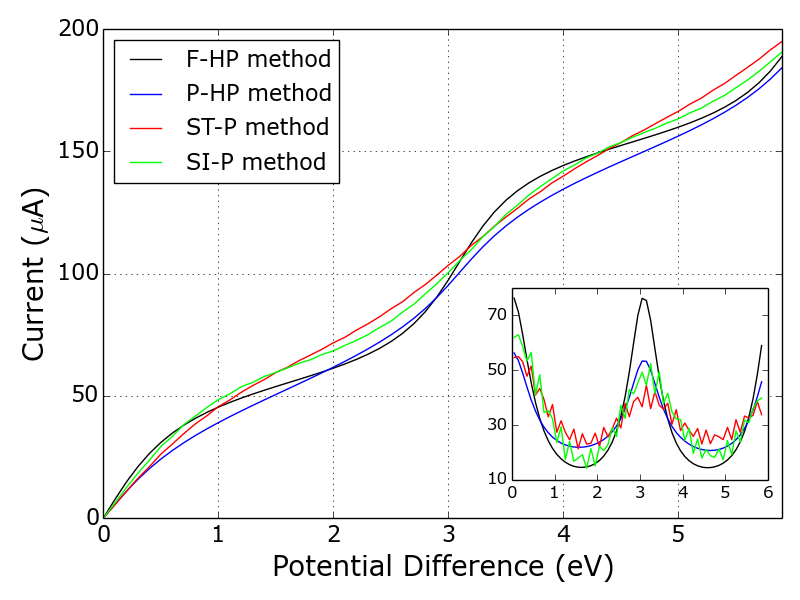}\\
\includegraphics[scale=0.42,keepaspectratio=true]{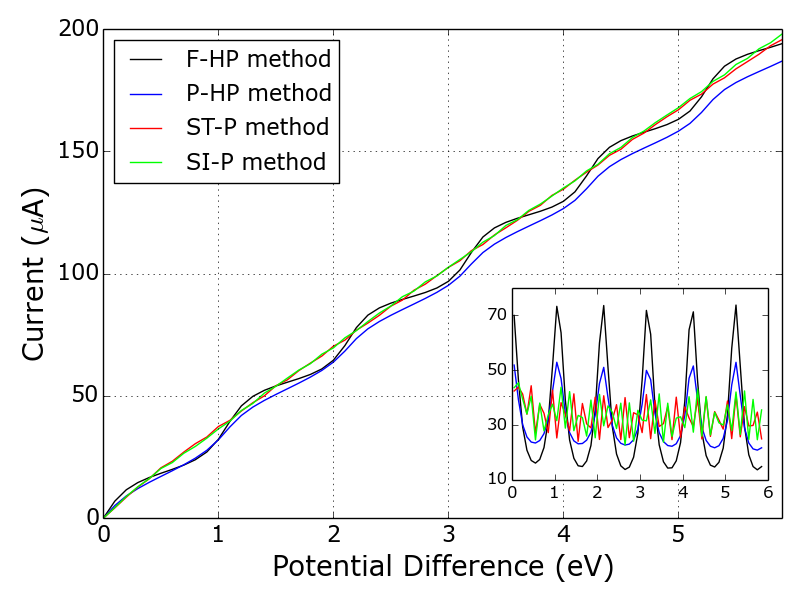}\\
\includegraphics[scale=0.42,keepaspectratio=true]{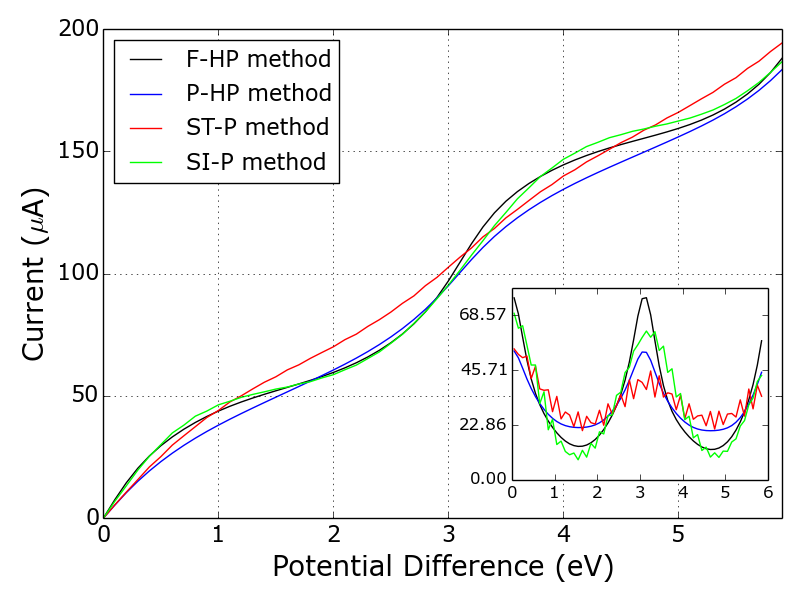}\\
\end{center}
\caption{Steady state currents as a function
of the applied potential, computed with different
implementations of the hairy-probes method with $\Gamma$ = 0.6 eV.
Results are shown for the resonant system with
different arrangements of the atoms in the central
region: 9-15-9 (top), 9-45-9 (center), and
30-15-30 (bottom).}
\label{IvsV-sigmoidal}
\end{figure}

As a matter of fact, the adoption of the sigmoidal field
proved to be at least as good as the step potential,
when not better, to reproduce the $I-V$ plots.
Figure \ref{IvsV-sigmoidal} presents these curves for
three resonant systems exhibiting different morphologies.
Aside from the original structure displayed in Figure
\ref{resonant-scheme}, two other models were examined
in which  the number of atoms in either the central or
the lateral segments was extended.
The value of $\Gamma$ was fixed to 0.6 eV, for which the
disagreement between methods was most noticeable.
For the standard resonant system the performances
obtained from  the step or the sigmoidal potentials
are comparable.
The same is true for the alternative resonant model
with a longer intermediate region, for which no
significant differences are found between the results
yielded by either method.
However, in this case the description provided by these
two approaches is manifestly worse.
In this sense, it is noteworthy that the P-HP scheme is
still able to capture the main features of the full
hairy probes curve.
On the other hand, in the third model where the device
bears longer lateral segments, the reference calculated
with the sigmoidal potential outperforms the one
calculated with the step-like field.

\section{The two different forms of the Driven Liouville-von Neumann formula}
\label{section.compare}

Equations \ref{master-tddft-z} and \ref{master-tddft-e}
are two expressions of the DLvN approach, both of which
had an heuristic origin, and for which different formal
derivation routes have been proposed.
From a mathematical point of view, the difference between
these two formulas is found in the off-diagonal elements
of the driving term: while in equation
\ref{master-tddft-z} these damp the lead-device
coherences to zero, in equation \ref{master-tddft-e}
coherences are pushed towards their equilibrium values.
Their derivations involve different approximations
and assumptions, and it is of particular interest
to compare the paths that lead to one or the other.
This is the goal of the present section, where the
performances of the two schemes are also confronted. 

It is possible to identify three assumptions or
approximations specific to one or the other formula,
that explain their different mathematical structures:

\begin{enumerate}
\item
In equation \ref{master-tddft-z}, the explicit lead
levels are in contact with an implicit fermionic
reservoir, for which coupling the wideband limit is
assumed.
Equation \ref{master-tddft-e} is in turn a truncation
of the hairy probes formula, where the leads are coupled
to a set of probes in which the wideband limit holds.
The latter procedure broadens but preserves the
electronic structure of the (finite) leads adjoining
the central region.

\item
To arrive to DLVN-z, it is assumed that the relaxation
dynamics in the leads is independent of the presence
of the device. This amounts to the zeroth-order
approximation of the Green's functions in which
$G^{adv}$ and $G^{ret}$ commute with the leads
subspace projector $Q$.
This step eventually leads to the disappearence of
$\rho^{eq}$ in the last term of equation 36 in the
work by Franco et al.\cite{franco}
Interestingly, the same result would also be obtained
e.g. from equations \ref{eqLC} or \ref{eqCL} of our
manuscript, if $\hat P_L$ or $\hat P_R$ were assumed
to commute with the Green's functions $\hat G^{\pm}_S$.
In such a case, these contributions would vanish,
suppressing $\rho^{0L}$ and $\rho^{0R}$ from the
off-diagonal elements of the equation of motion,
and thus giving rise to a driving term analogous
to DLvN-z.
Thus, the damping of the coherences either to zero or
to equilibrium is related to this approximation in the
Green's functions, which physically translates into an
independency of the electron  dynamics in the leads
with respect to the device.

\item
Finally, the $\hat \Omega$ terms are dropped from the
HP expression to arrive at DLvN-e.
These terms arise from the Green's functions keeping
track of the forward time-propagation of carriers
coming from the probes.
Thus, they introduce a propagation sense to the injected
lead-device charge. Its suppression must imply a drop in
the current, as is certainly observed in the simulations.

\end{enumerate}

\begin{figure}
\begin{center}
\includegraphics[scale=0.40,keepaspectratio=true]
{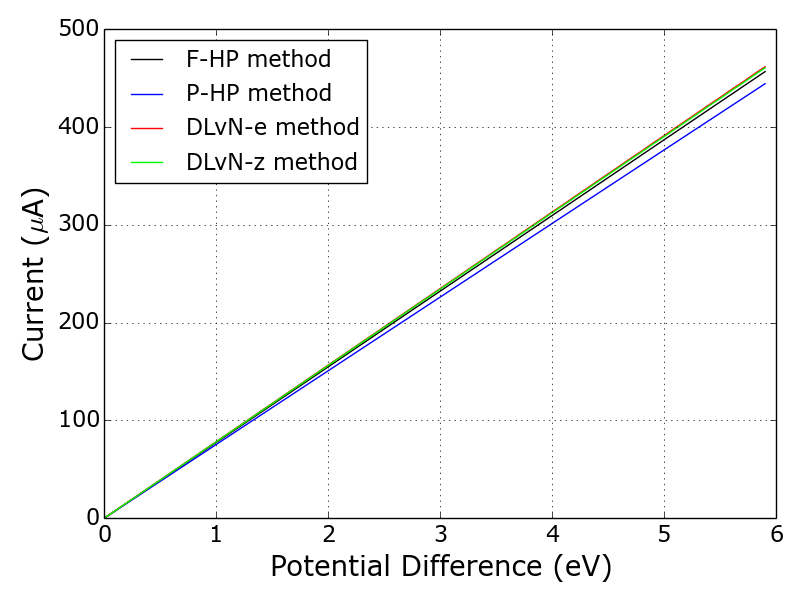}
\includegraphics[scale=0.40,keepaspectratio=true]
{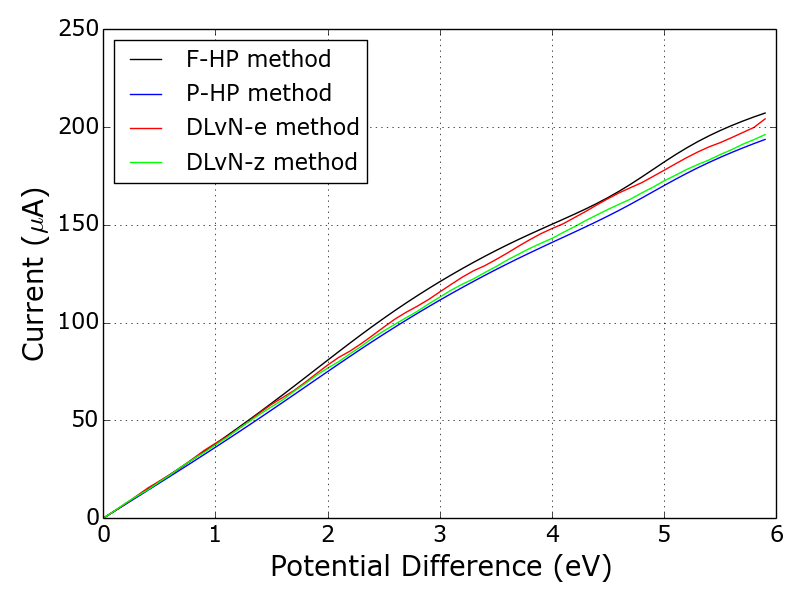} \\
\includegraphics[scale=0.40,keepaspectratio=true]
{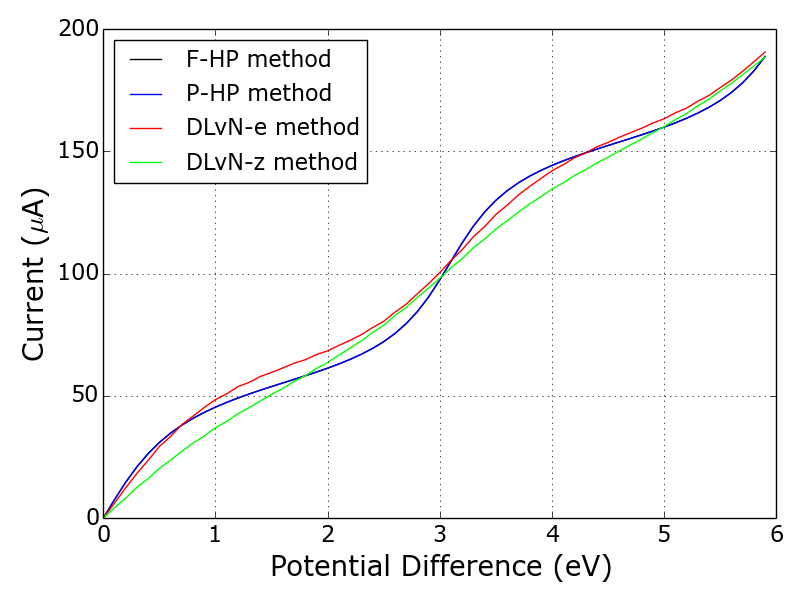}
\includegraphics[scale=0.40,keepaspectratio=true]
{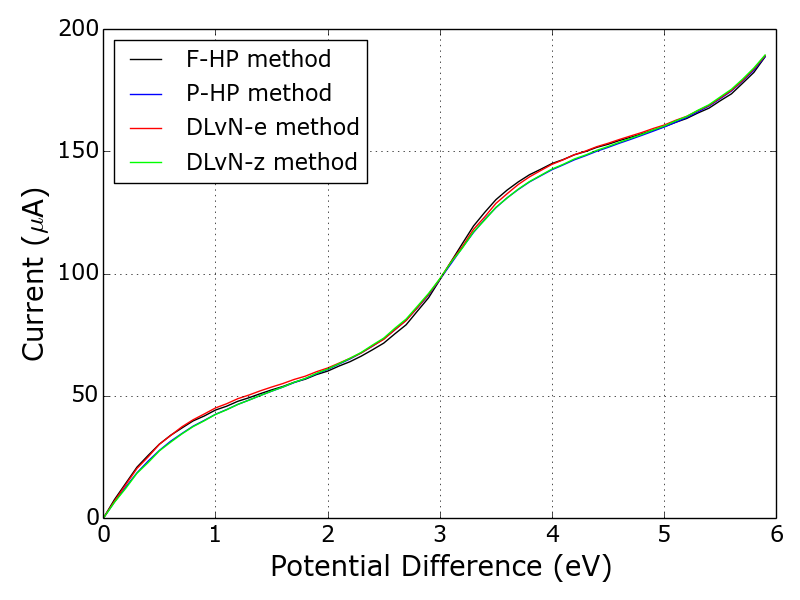}
\end{center}

\caption{Steady state current as a function of the
applied bias computed with the two forms of the DLvN
equation and with the hairy-probes method.
The top panel presents the data corresponding to the
ballistic (left) and disordered (right) systems, for
$\Gamma$ equal to 0.6 eV.
The bottom panel shows the results for the resonant
system with $\Gamma$ equal to 0.6 eV (left) and
0.1 eV (right).}
\label{confront1}
\end{figure}

In ref. \cite{jcp_morzan} ~ it has been reasoned that
while DLvN-z represents a device between electronic
reservoirs at equilibrium with well-defined chemical
potentials, DLvN-e resembles the state of a charged
capacitor, where the target density represents the
equilibrium state of the entire finite junction model
and not just the leads. This seems to be consistent
with the origins of each of these schemes, that have
now come to light.
Figure \ref{confront1} depicts the current-voltage
curves for the two DLvN approaches, together with the
results of the HP method. At small couplings it can be
seen that the performance of any of these methods is
essentially indistinguishable from each other.
As the $\Gamma$ parameter is increased, discrepancies
start to emerge.

\begin{figure}
\begin{center}
\includegraphics[scale=0.06,keepaspectratio=true]{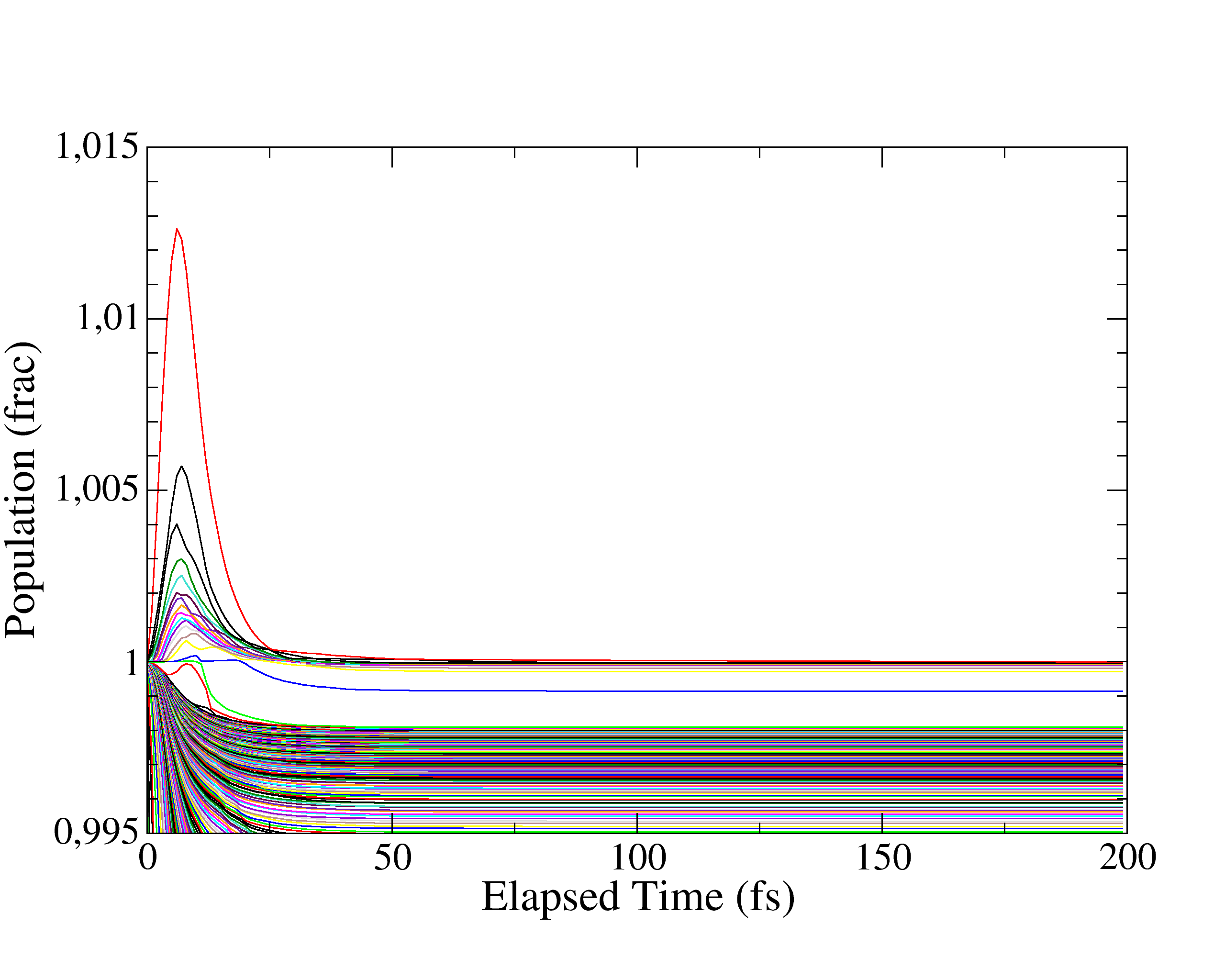}
\includegraphics[scale=0.06,keepaspectratio=true]{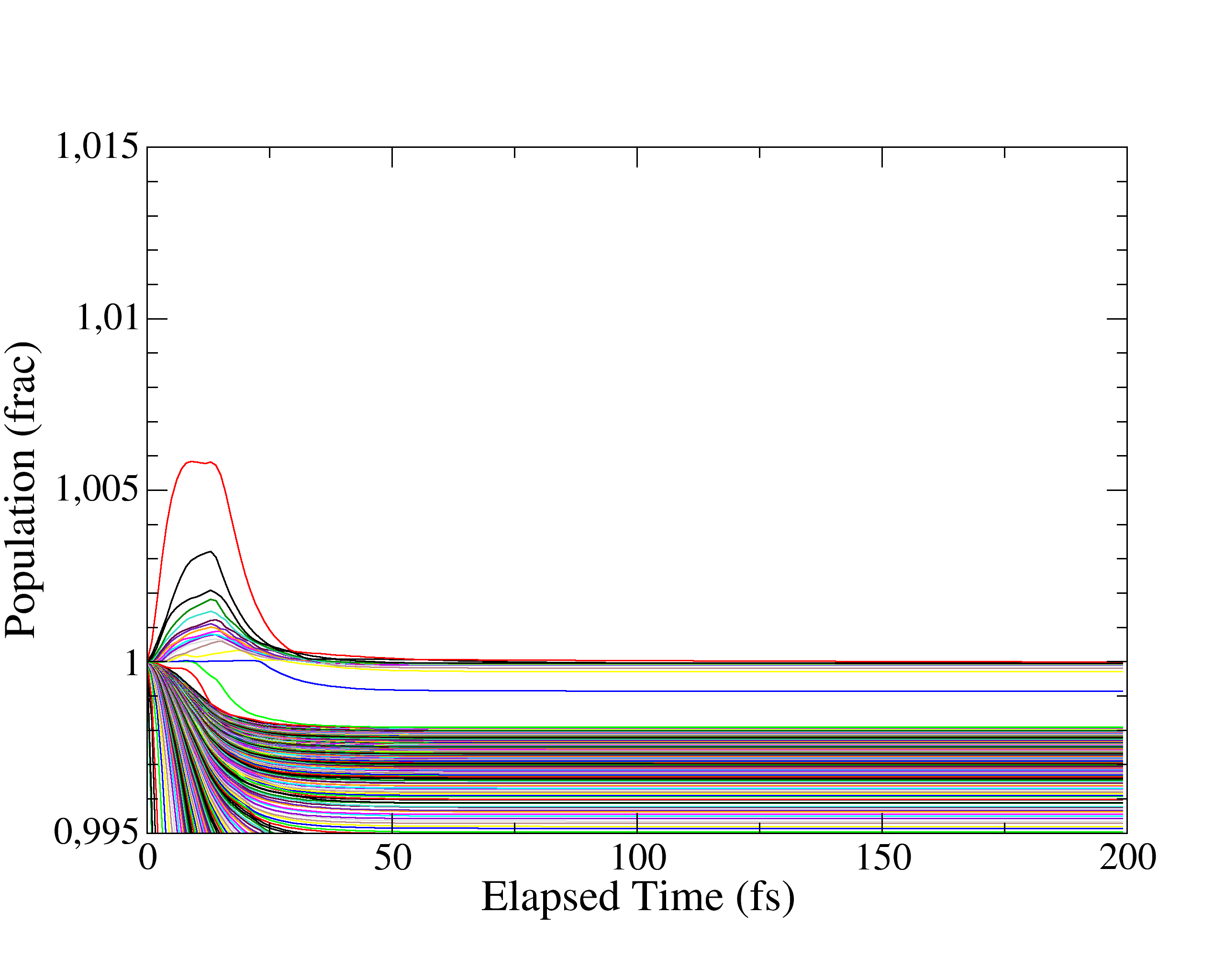}
\includegraphics[scale=0.06,keepaspectratio=true]{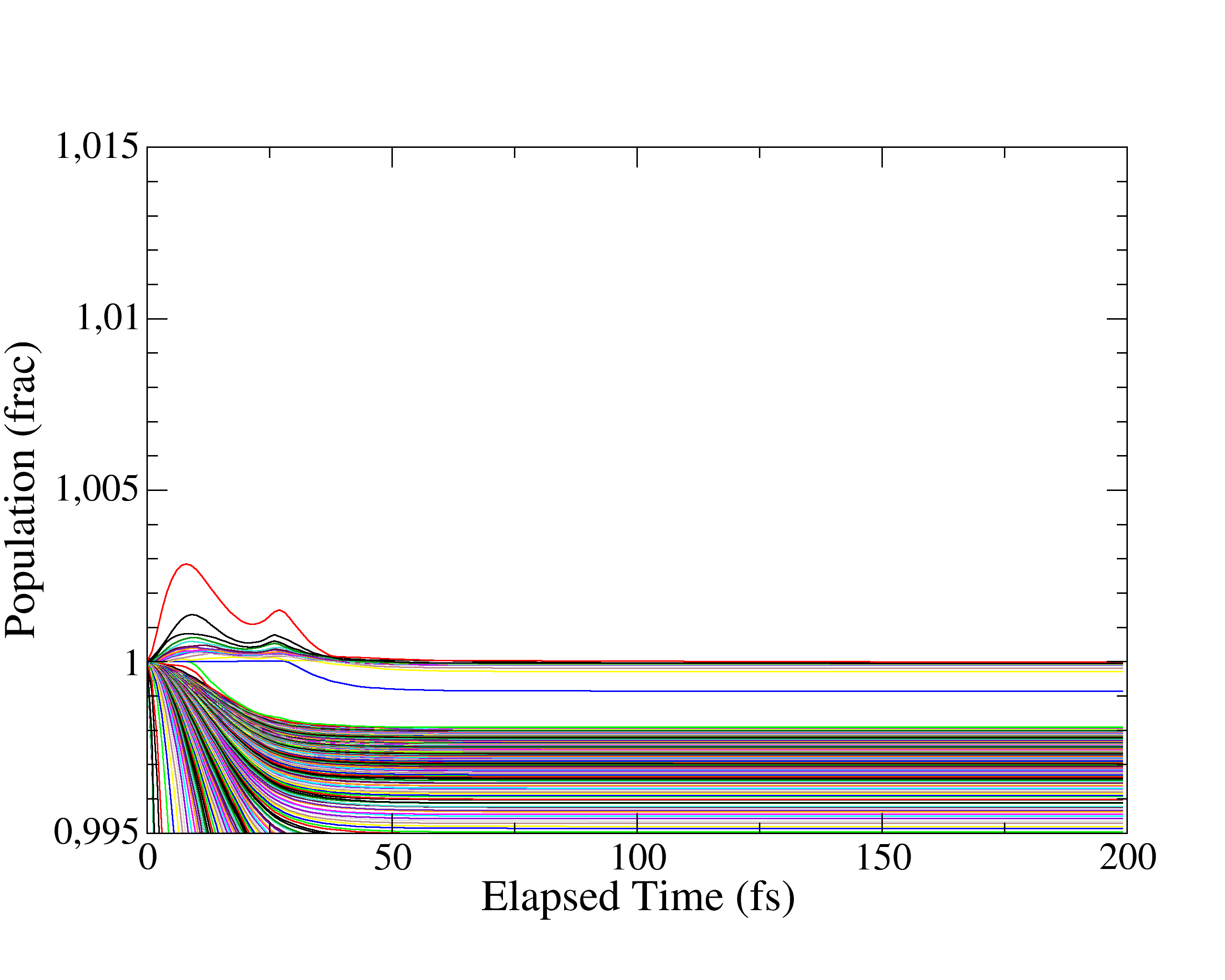} \\
\includegraphics[scale=0.06,keepaspectratio=true]{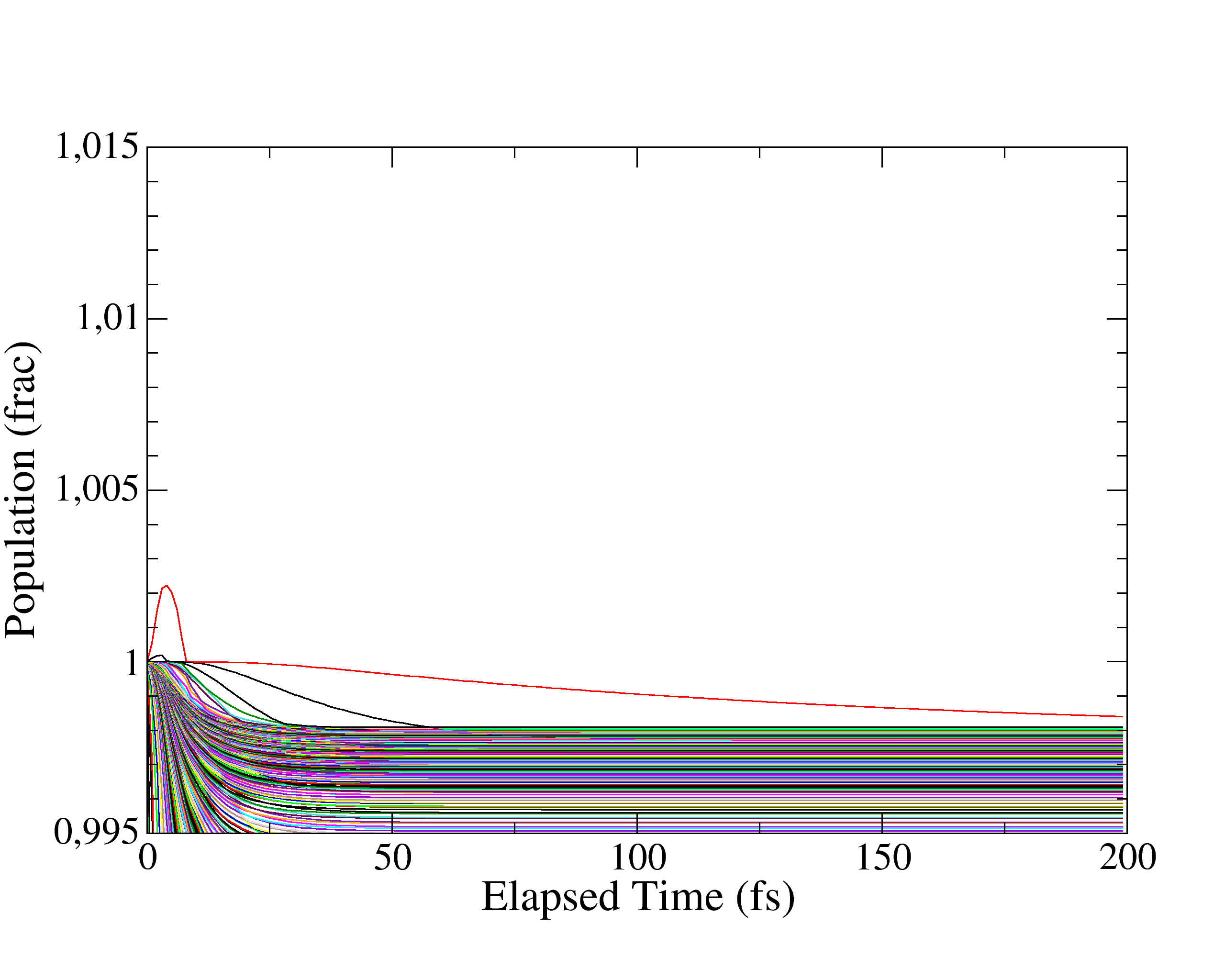}
\includegraphics[scale=0.06,keepaspectratio=true]{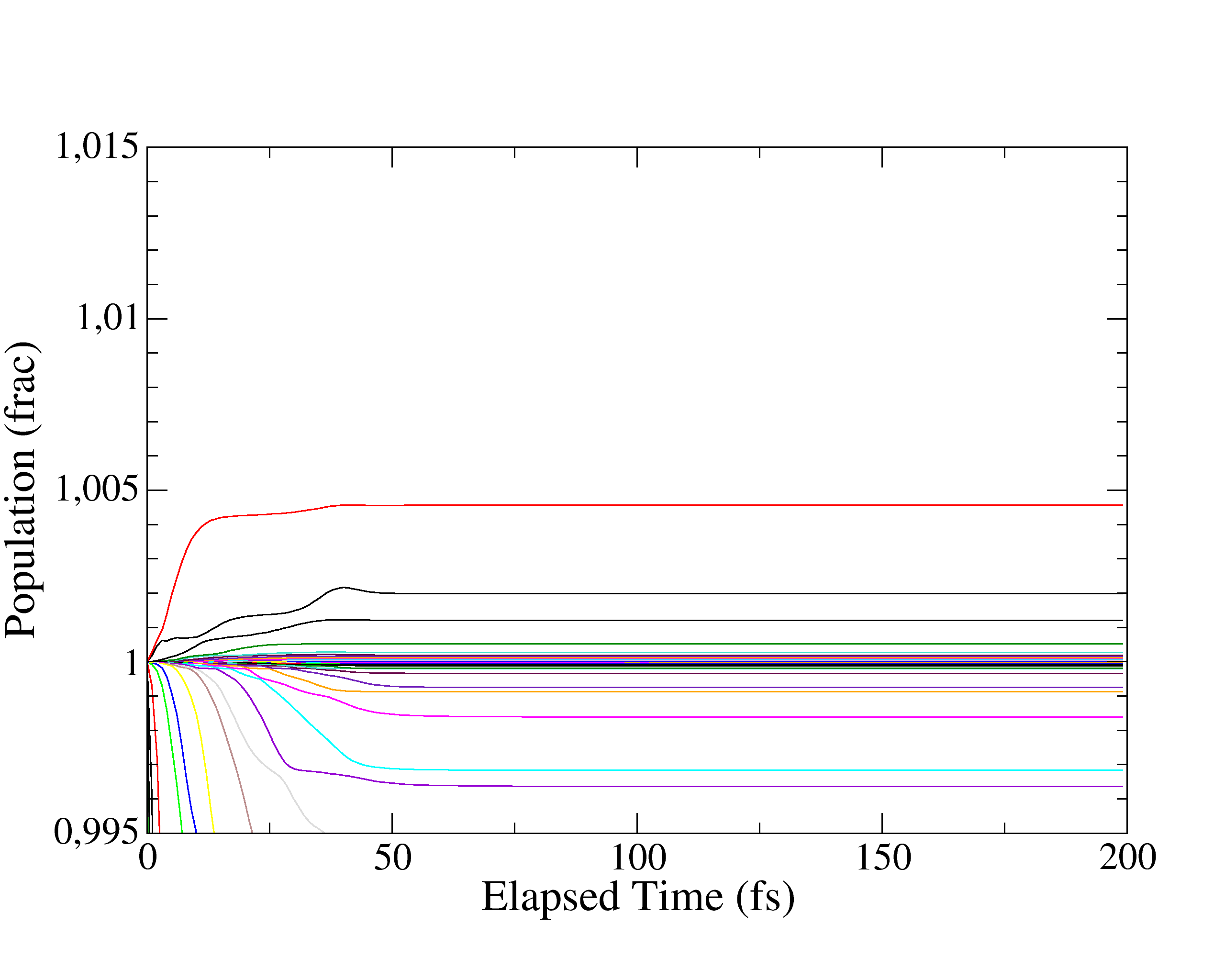}
\includegraphics[scale=0.06,keepaspectratio=true]{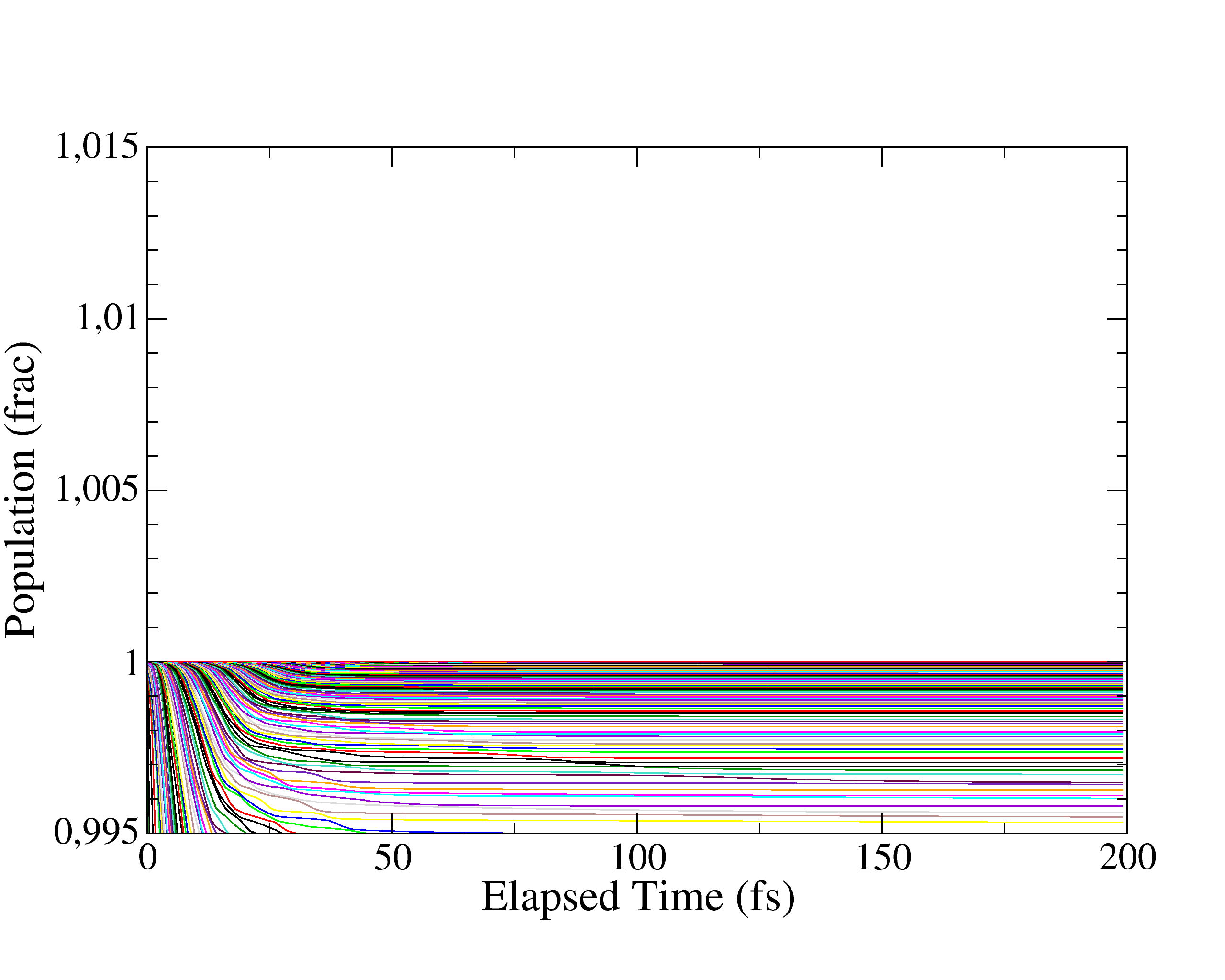}
\end{center}

\caption{Temporal evolution of the eigenvalues of
the density matrix for the resonant model system.
The top row corresponds to the full HP method where
the driving term was appplied gradually using ramps
of different durations: 5.0 fs, 12.5 fs and 25.0 fs
(from left to right).
The bottom row compares the behavior found with the
other methods using a ramp of 5.0 fs: the P-HP (left),
the DLvN-e (center) and the DLvN-z (right) schemes.
Part of this data has been already given in a
different size scale in Figure
\ref{eigenstatespop}.}
\label{pauli}
\end{figure}

DLvN-z has been shown to observe Pauli's principle
regardless of the initial conditions.\cite{jctc_10_2927}
It is interesting to note that, while the hairy probes
scheme also fulfills Pauli's principle at long times
(by construction), it does not necessarily obey it
in the transient.
This can be seen in Figure \ref{pauli}, which displays
the occupations for F-HP, P-HP, DLvN-e, and DLvN-z in
the resonant system.
It must be recalled that in all these schemes, the
dynamics is switched on smoothly, to avoid sudden jumps
which could lead to numerical discontinuities.
Specifically, the driving term and the $\Omega$ term are
introduced in the first part of the simulation using a
linear ramp.
For F-HP, the positive and negative deviations from
1 and 0 respectively---which become smaller with a
smoother ramp---disappear in the long term.
When the $\Omega$ term is suppressed from the F-HP scheme,
the dynamics and in particular the occupations are affected
but the exclusion principle is still obeyed in the steady
state.
However, if the reference density matrix is replaced
by the one obtained from an electric field to give
the DLvN-e approach, the violation to the exclusion
principle persists even in the steady state.
This is not observed with the DLvN-z method, for which
the exclusion principle is satisfied throughout the full
dynamics, despite the adoption of the same $\hat \rho^0$
as in DLvN-e.
Interestingly, this makes manifest that the observance of
Pauli's principle is determined neither by $\Omega$ nor
by the reference density, but by their combination.

\begin{figure}
\begin{center}
\includegraphics[scale=0.40,keepaspectratio=true]
{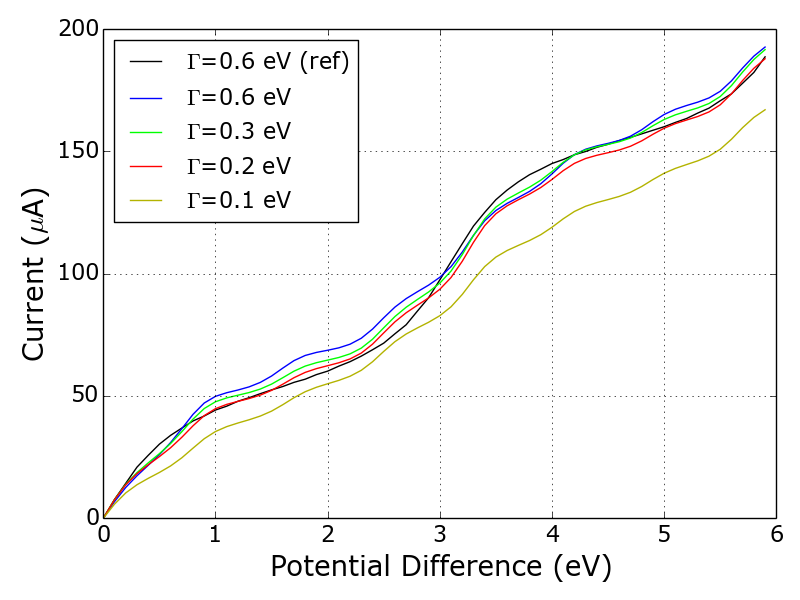}
\includegraphics[scale=0.40,keepaspectratio=true]
{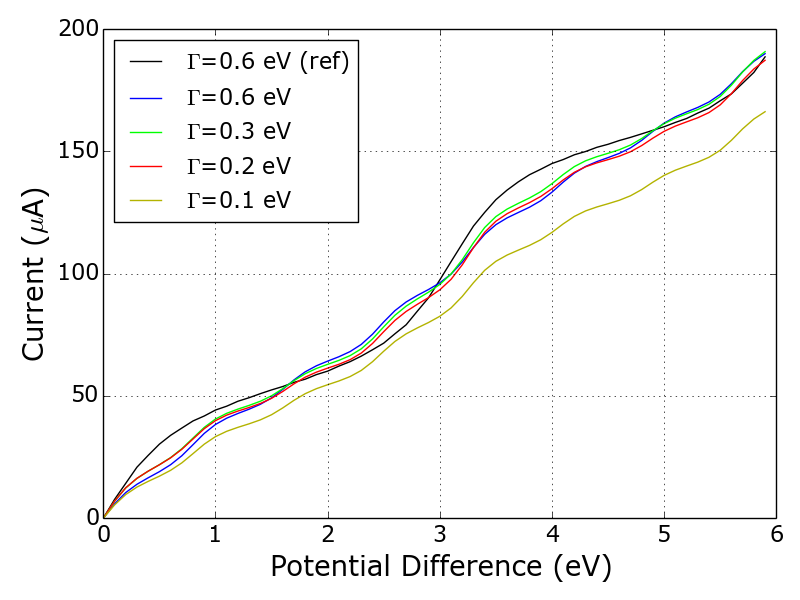} \\
\includegraphics[scale=0.40,keepaspectratio=true]
{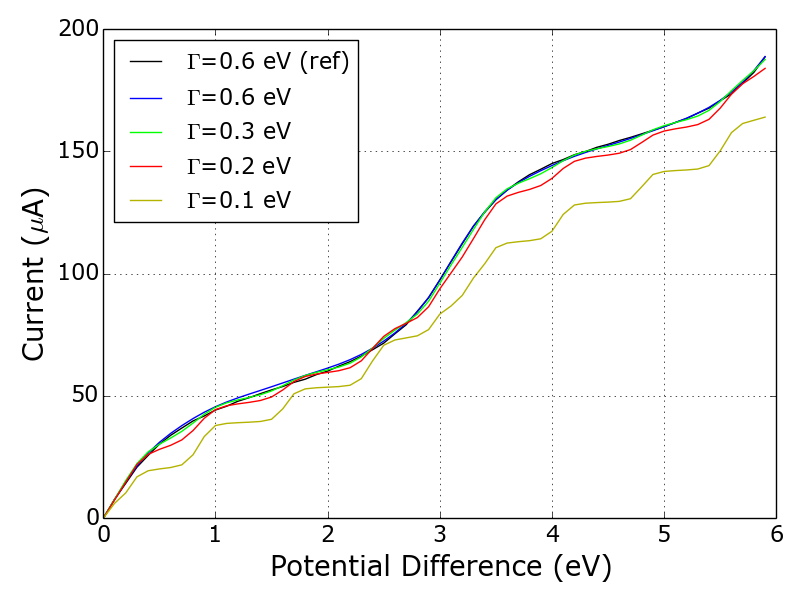}
\includegraphics[scale=0.40,keepaspectratio=true]
{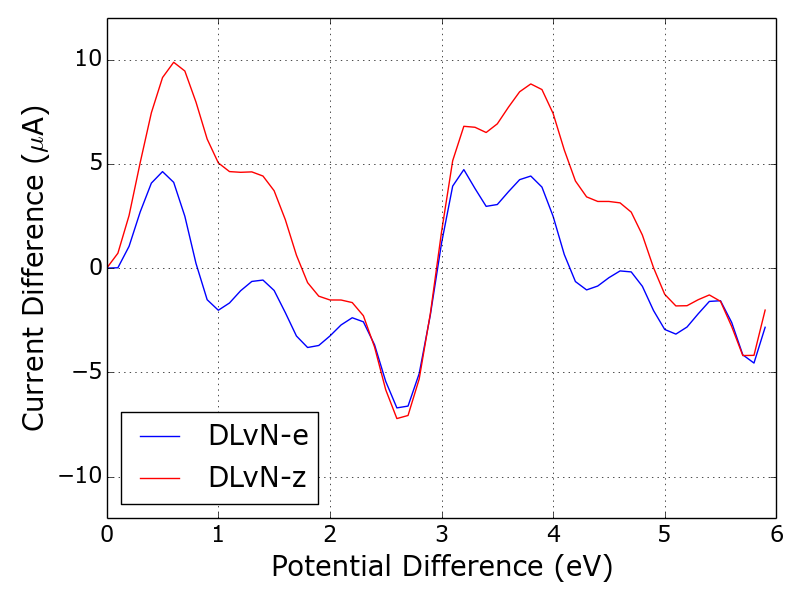}
\end{center}
\caption{Effect of electrode size and of the $\Gamma$
parameter on the current-voltage curves obtained for
the resonant model.
Top panels present in black the curve obtained from F-HP
with 200 atoms in each lead and $\Gamma$=0.6 eV
(labelled ref), in comparison with the curves computed
using 50 atoms leads and different $\Gamma$ values for
the DLvN-e (left) and DLvN-z (right) methods.
We also include the behaviour of the F-HP method with
the small electrodes and varying $\Gamma$ (bottom left).
The bottom right panel displays, for DLvN-e and DLvN-z ($\Gamma$ = 0.3 eV),
the difference with respect to the F-HP current (obtained with $\Gamma$ = 0.6 eV).}
\label{fig-size}
\end{figure}

Finally, Figure \ref{fig-size} explores the effect of
$\Gamma$ and of the electrode size on the currents,
using the resonant model.
The bottom left panel shows that the HP method,
consistently with its physical origin, is particularly
robust with respect to electrode size and $\Gamma$.
In the case of an electrode of 50 atoms, significant
deviations in the current are observed only for
$\Gamma<$0.2 eV.
The  two upper panels depict the I-V curves for the
DLvN-e and DLvN-z approaches, compared with the F-HP
method.
In both cases it can be observed that the reduction of
the electrode size to 50 atoms affects the currents.
The agreement with F-HP improves marginally as $\Gamma$
decreases, until it breaks down when this parameter
falls below the wideband limit, which in this case
is higher since the energy level spacing becomes
larger with smaller leads. 
This distortion in the currents is less critical
in the DLvN-e approach.
In particular, the bottom right panel displays the
difference between the benchmark F-HP current
($\Gamma$=0.6 eV, 200 atoms in the leads) and the
currents produced by the DLvN approaches
($\Gamma$=0.3 and 50 atoms in the leads).
It can be seen that deviations tend to be larger
for DLvN-z, whereas the DLvN-e approach, inheriting
the mathematical structure from F-HP, copes better
with the shortening of the electrodes.

To summarize, the two forms of the Driven Liouville-von
Neumann equation produce similar results, reproducing
the hairy probes method provided large values of $\Gamma$
are avoided.
The most noticeable difference is that, whereas DLvN-z
observes Pauli's principle, DLvN-e does not.
At the same time, in some situations the DLvN-e equation
is more tolerant to a decrease of electrode size, as
discussed in ref.\cite{jcp_morzan} ~, which is a
consequence of the robustness of the HP method from
which it is descended.

\section{Summary and final remarks}
\label{section.conclus}

In this article, it was shown that the Green's functions
based multiple-probes---or hairy probes---formalism, 
adopts a form equivalent to the heuristic Driven
Liouville-von Neumann method as proposed in reference
~\cite{jcp_morzan}  (equation \ref{master-tddft-e}),
plus an additional term involving a matrix ($\hat \Omega$)
with null diagonal blocks.
A distinctive feature of this form of the DLvN equation
of motion is that, at variance with the previous versions
introduced in references ~\cite{jcp_124_214708}~ or ~\cite{jctc_10_2927}~, the coherences are damped to
the equilibrium density.
It has been argued that in this approach the electrodes
are not meant to represent infinite reservoirs with
homogeneous and well defined chemical potentials.
Instead, through the action of the driving operator,
the leads are driven close to, rather than exactly to,
the target density.\cite{jcp_morzan}
This is parallel to the physics in the HP model, where
the electrodes are connected to multi-probes with
electrochemical potentials $\mu_{L/R}$.
The electrochemical potential of the lead is not
necessarily that of the probes but depends on the
strength of their coupling (and on position down
the lead).
These particular boundary conditions of the HP method
are reminiscent of those in the DLvN implementation of
equation \ref{master-tddft-e} and demonstrated to be
well suited for small size electrodes.\cite{jcp_morzan}

To neglect the $\hat \Omega$ matrix in the driving
term of the HP formula has minor effects on the
dynamics and the steady state currents obtained
for a variety of model systems. 
These effects are even less significant when the
coupling between the probes and the leads is reduced
by decreasing the $\Gamma$ parameter.
More specifically, our results show that the P-HP method
can reproduce the behavior described by the full HP
scheme for ballistic and disordered systems, and with
some care it can also be tuned for more complicated
resonant systems.
In general, at least in the context of tight-binding
Hamiltonians, it is possible to conclude that the DLvN
method, incarnated here in the P-HP or ST-P equations
of motion, converges to the hairy probes description
in the limit of small couplings between the probes and
the leads (providing $\Gamma$ is still larger than the
energy level spacing). 
The P-HP method can be thought of as a version of the
DLvN approach in which the calculation of the reference
density is based  on Green's functions.
The ST-P method, on the other hand, reproduces the
formulation presented in reference ~\cite{jcp_morzan}~.
It has proved to be a very good approximation to the
full HP description, but the finite size effects which
manifest in the absence of a self-energy imply a
limitation in comparison with the P-HP approach that
seems difficult to overcome.

The possibility to avoid the calculation of Green's
functions acquires special interest for the applications
in the context of TDDFT or other first-principles schemes.
In that respect, we explored the substitution of the
reference density computed from Green's functions in the
P-HP method, by the one emerging from an electric field
applied to the system. 
This strategy systematically produced higher currents
than the P-HP method, presumably because the application
of a field, in the present setting, amounts to an additive
constant in the density matrix elements, whereas in the HP
scheme the chemical potential is fixed at an external
probe and not directly on the leads, whose polarization
is mediated by a coupling parameter.
The effective bias arising between the electrodes in
operation conditions may thus not be as large as the
one imposed by the field.
Further sources of improvement for the region-weighted
field-generated reference density method proved hard
to find.
Our results here seem to suggest that the step and sigmoideal shapes fields
can better fit those parts of the reference matrices at the boundary with the device,
providing the most  accurate representations of the current in comparison with the 
other fields tested. Whereas the ballistic and disordered models could be described reasonably
well with the ST-P and SI-P schemes, the resonant systems proved to be challenging for those
methods using electric-field based reference densities.
This limitation becomes particularly relevant for
TDDFT applications, where the complexity of the
chemical structures, far from the simple tight-binding
models examined in the present work, may lead to stronger
discrepancies between the results obtained using a
reference density generated from Green's functions or
from an electric field.
It would be interesting to establish how the different
reference densities and the omission of the
$\hat \Omega$ matrix affect the dynamics in the case
of first-principles simulations.
That question will be the subject of future work.

\begin{acknowledgement}
This work was supported by a research grant from
Science Foundation Ireland (SFI) and the Department
for the Economy Northern Ireland under the SFI-DfE
Investigators Programme Partnership, Grant Number
15/IA/3160, by the Argentinean Agency for Scientific
and Technological Promotion (ANPCYT) through PICT
2015-2761, and by  the University of Buenos Aires,
UBACYT 20020160100124BA.
We are grateful to Uriel Morzan for very appreciated
discussions.
\end{acknowledgement}

\providecommand{\latin}[1]{#1}
\providecommand*\mcitethebibliography{\thebibliography}
\csname @ifundefined\endcsname{endmcitethebibliography}
  {\let\endmcitethebibliography\endthebibliography}{}

\end{document}